\documentclass{aa}
\usepackage{graphicx}
\usepackage{natbib}
\usepackage[varg]{txfonts}
\usepackage{amssymb}

\usepackage{multirow}

\def\apjl{ApJL}%
\def\aap{A\&A}%
\def\apjs{ApJS}%
\begin{document}

\title{The Lyman Alpha Reference Sample IX: Revelations from deep surface photometry\thanks{Based on observations made with the Nordic Optical Telescope, operated by the Nordic Optical Telescope Scientific Association at the Observatorio del Roque de los Muchachos, La Palma, Spain, of the Instituto de Astrofisica de Canarias. Based [in part] on observations obtained with WIRCam, a joint project of CFHT, the Academia Sinica Institute of Astronomy and Astrophysics (ASIAA) in Taiwan, the Korea Astronomy and Space Science Institute (KASI) in Korea, Canada, France, and the Canada-France-Hawaii Telescope (CFHT) which is operated by the National Research Council (NRC) of Canada, the Institut National des Sciences de l'Univers of the Centre National de la Recherche Scientifique of France, and the University of Hawaii. Based [in part] on observations made with the NASA/ESA Hubble Space Telescope, obtained at the Space Telescope Science Institute, which is operated by the Association of Universities for Research in Astronomy, Inc., under NASA contract NAS 5-26555. These observations are associated with program \#12310,\#11522.} }
\author{Genoveva Micheva\inst{1}
\and G\"oran \"Ostlin\inst{2} 
\and Erik Zackrisson\inst{3}
\and Matthew Hayes\inst{2}
\and Jens Melinder\inst{2}
\and Lucia Guaita\inst{4}
\and John M. Cannon\inst{5}
\and Joanna S. Bridge\inst{6}
\and Daniel Kunth\inst{7}
\and Andreas Sandberg\inst{2}
}
\institute{University of Michigan, 311 West Hall, 1085 S. University Ave, Ann Arbor, MI 48109-1107, USA
\and
Stockholm Observatory, Department of Astronomy, Stockholm University, 106\,91 Stockholm, Sweden
\and
Department of Physics and Astronomy, Uppsala University, Box 515, SE-751 20 Uppsala, Sweden
\and
N\'ucleo de Astronom\'ia, Facultad de Ingenier\'ia, Universidad Diego Portales, Av. Ej\'ercito 441, Santiago, Chile
\and
Macalester College, 1600 Grand Avenue, Saint Paul, MN 55105, USA
\and
Department of Physics and Astronomy, 102 Natural Science Building, University of Louisville, Louisville, KY 40292, USA
\and
Institut d'Astrophysique de Paris, 98 bis Bd Arago, F-75014 Paris, France
}

\date{Received ...; Accepted 16/03/2018}


\abstract
{The Lyman $\alpha$ Reference Sample (LARS) of $14$ star-forming galaxies offers a wealth of insight into the workings of these local analogs to high-redshift star-forming galaxies. The sample has been well-studied in terms of Ly$\alpha$ and other emission line properties, such as H I mass, gas kinematics, and morphology.}{We analyze deep surface photometry of the LARS sample in $UBIK$ broadband imaging obtained at the Nordic Optical Telescope and the Canada-France-Hawaii Telescope, and juxtaposition their derived properties with a sample of local high-redshift galaxy analogs, namely, with blue compact galaxies (BCGs). }{We construct radial surface brightness and color profiles with both elliptical and isophotal integration, as well as RGB images, deep contours, color maps, a burst fraction estimate, and a radial mass-to-light ratio profile for each LARS galaxy. Standard morphological parameters like asymmetry, clumpiness, the Gini and M20 coefficients are also obtained and analyzed, as well as isophotal asymmetry profiles for each galaxy. In this context, we compare the LARS sample to the properties of the BCG sample and highlight the differences.}{Several of these diagnostics indicate that the LARS galaxies have highly disturbed morphologies even at the level of the faintest outer isophotes, with no hint at a regular underlying population, as found in many BCG sample galaxies. The ground-based photometry reaches isophotes down to $\sim28$ mag arcsec$^{-2}$, while the space-based data reach only $\sim26$ mag arcsec$^{-2}$. The ground-based observations therefore reveal previously unexplored isophotes of the LARS galaxies. The burst fraction estimate suggests a spatially more extended burst region in LARS than in the BCGs. Comparison to stellar evolutionary models in color-color diagrams reveals complex behavior of the radial color profiles, often inconsistent with a single stellar population of any age and metallicity, but instead suggesting a mixture of at least two stellar populations with a typical young mass fraction in the range $0.1$-$1\%$. }{The galaxies in the LARS sample appear to be in earlier stages of a merger event compared to the BCGs. Standard morphological diagnostics like asymmetry, clumpiness, Gini and M20 coefficients cannot separate the two samples, although an isophotal asymmetry profile successfully captures the average difference in morphology. These morphological diagnostics do not show any correlation with the equivalent width or the escape fraction of Lyman Alpha. }


\titlerunning{LARS deep surface photometry}
\authorrunning{Micheva et al.}

\maketitle

\section{Introduction}
The Lyman $\alpha$ Reference Sample (LARS) is a sample of $14$ star-forming galaxies at redshifts $0.028<z<0.18$. In a series of recent works, the properties of this sample have been thoroughly studied. In \citet{Ostlin2014} the sample is presented in H$\alpha$, H$\beta$ narrowband, and u, b, i broadband imaging with the Hubble Space Telescope's ({\it HST}) Advanced Camera for Surveys (ACS) and the Wide Field Camera 3 (WFC3). This work illustrates the potential of the LARS to be used as a sample of local analogs reminiscent of high-redshift star-forming galaxies, like Lyman alpha emitters (LAEs) and Lyman break galaxies (LBGs). The properties and the geometry of the Ly$\alpha$ line are investigated and found to be bright, extended, and strongly asymmetric for the test case of LARS01, which has a high escape fraction of Ly$\alpha$ of $f_{esc}^{{Ly}\alpha }=12\%$. \citet{Hayes2014} study the intensity of LARS galaxies in Ly$\alpha$, H$\alpha$, and UV, and present Balmer decrement maps and equivalent widths of Ly$\alpha$. They also investigate the radial Ly$\alpha$ and UV continuum profiles of the galaxies. The Ly$\alpha$ emission is well-fitted by exponential disk-like profiles, with Sersic indices of $n=1\mbox{-}2$, while the UV continuum is more consistent with de Vaucouleurs $n=4$ and increasingly convex ($n>4$) profiles. These latter authors further find higher Ly$\alpha$ escape fractions in the galaxies with lower mass, star formation rates (SFR), dust content and metallicities. Their work demonstrates that the LARS sample contains low-redshift Ly$\alpha$ emitters with some of the highest escape fractions and equivalent widths, with one object (LARS02) showing $f_{esc}^{{Ly}\alpha }=80\%$ - rare at any redshift. Such high escape fractions are similar to those found at high redshifts $(z\ge3)$. These authors show that the LARS sample is analogous in luminosity to high-redshift LAEs and LBGs. They are also bright enough in FUV absolute magnitude to be detectable in high-redshift LAE and LBG surveys.\\

\noindent \citet{Pardy2014} provide H I masses for the individual LARS galaxies, except the three highest-redshift ones. The observations reveal complex tidal structures and disturbed velocity fields in some of the LARS galaxies, hinting at major merger events in the recent evolution of these systems. \citet{Guaita2015} investigate the effect that redshifting the LARS sample to higher redshifts of $z\sim2$ and $z\sim5.7$ would have on morphological parameters such as asymmetry, concentration, Gini, and M20 coefficients. These authors find the morphology is consistent with mergers, and that the Ly$\alpha$ escape seems uncorrelated with the observed morphology and morphological parameters. \citet{RiveraThorsen2015} use high-resolution far-UV {\it HST} spectroscopy to study metal absorption lines, probing the neutral interstellar medium (ISM), and estimate optical depths and covering fractions of the gas. They find outflows for the majority of LARS, and show that the presence and strength of the outflows does not guarantee strong Ly$\alpha$ emission. \citet{Duval2016} find two bipolar outflowing halos of neutral gas, seemingly associated with regions bright in Ly$\alpha$, in integral field unit (IFU) H$\alpha$ data of LARS05. Their radiative transfer modeling successfully reproduces observed emission line ratios, including the shape and strength of the Ly$\alpha$ line. These authors suggest that such outflows may help explain the strong Ly$\alpha$ emission observed in high-redshift galaxies. \citet{Herenz2016} use IFU spectroscopy to study the H$\alpha$ line kinematics and find that some of the LARS galaxies appear to be dispersion-dominated, with high-velocity-dispersion galaxies showing a higher escape fraction of Ly$\alpha$. Such a turbulent ISM is likely similar to the conditions found at high redshift. \citet{Bridge2018} examine the dust geometry in the LARS sample and find it to be clumpy for most galaxies.\\

\noindent Large amounts of multiwavelength data have been amassed for these studies in an effort to provide as complete a coverage as possible of these strong Ly$\alpha$ emitters, including that from the Hubble Space Telescope ({\it HST}), Green Bank Telescope (GBT), Very Large Array (VLA), Potsdam Multi-Aperture Spectrophotometer (PMAS), Nordic Optical Telescope (NOT), and the Canada-France-Hawaii Telescope (CFHT). The LARS galaxies thus represent excellent local laboratories in which to study conditions similar to high-redshift galaxies, but being nearby, they conveniently lend themselves to high-resolution analysis, impossible at high redshift. \\

\noindent Another class of nearby high-redshift analogs are the blue compact galaxies (BCGs). They are an inhomogeneous group with wide ranges of all of their properties. Their metallicities range from the lowest ever observed in a galaxy \citep[][SBS 0335-052W with $12+\log{O/H}=7.12\pm0.03$, and J0811+4730 with $12 + \log{O/H} = 6.98 \pm 0.02$, respectively]{Izotov2005,Izotov2018}, to solar \citep{Masegosa1994}. Their star formation rates (SFRs) span approximately five orders of magnitude up to tens of solar masses per year \citep{Hopkins2002}. Some BCGs show a nuclear starburst in an otherwise red, old, and extended stellar population, while others display highly irregular morphologies with numerous starbursting regions \citep[e.g.,][]{Loose1986,Kunth1986,Salzer1989,Papaderos1996,Telles1997, Cairos2001,Bergvall2002, Micheva2013a,Micheva2013b}. While BCGs constitute a very diverse group, the presence of a starburst taking place in an underlying host galaxy is the major commonality among them.\\

\noindent In this paper, we investigate what an analysis of deep surface photometry in the optical and near infrared (NIR) of the LARS galaxies can add to their already thorough characterization. Specifically, with our deep ground-based data we can reach a previously unexplored isophotal range in LARS, namely the $26$-$28$ mag arcsec$^{-2}$ region. We attempt to isolate and study separately the properties of the current starburst and the underlying older population, and estimate the burst fraction and the stellar mass contained in, for example, the $26$-$28$ mag arcsec$^{-2}$ region. We also investigate how the properties we derive in this work compare to the other well-known group of high-redshift analogs, the BCGs, and search for ways to separate the two samples using standard morphological diagnostic tools. We perform this analysis in the context of the previously established parameter space in \citet{Micheva2013a} and \citet{Micheva2013b}. The paper is structured as follows: In Section ~\ref{sec:data} we present the observational data used in the analysis. All diagnostic tools, like surface brightness and color profiles, deep contour and color maps, color-color diagrams, and so on, are presented in Section~\ref{sec:tools}. The implications of our analysis and the comparison to BCGs can be found in Section \ref{sec:discuss}, and our conclusions are summarized in Section \ref{sec:conclusions}. We highlight individual characteristics of each LARS galaxy in the appendix, where we also present the analysis figures for LARS02 through LARS14.\\

\noindent Throughout this paper we use the AB magnitude system and assume a flat $\Lambda$CDM cosmology with $H_{0}=70$km s${}^{-1}$ Mpc${}^{-1}$, $\Omega_{\textrm{M}}=0.3$, and $\Omega_{\Lambda}=0.7$. The BCG photometry from \citet{Micheva2013a,Micheva2013b} was converted from Vega mag to AB mag to fascilitate the comparison in this paper. All surface brightness values are presented in AB mag arcsec$^{-2}$.

\begin{table}
\caption{Exposure times for each galaxy and filter. LARS06 was observed with ALFOSC instead of MOSCA. {\it HST}/WFC3 data are indicated by a $\dagger$.}
\label{tab:obslog}
\centering
\begin{tabular}{c c c c c}
\hline\hline                
LARS  & $U$  & $B$ & $I$ & $K$ \\   
\hline                       
   01 &  2$\times$900s  &  7$\times$300s  &  8$\times$200s  & 333$\times$20s \\
   02 &  4$\times$900s  & 17$\times$300s  &  8$\times$200s  & 330$\times$20s \\
   03 &                 & 6$\times$300s   &                 & 150$\times$20s \\
   04 &  2$\times$900s  & 11$\times$300s  &  8$\times$200s  & 267$\times$20s \\
   05 &  $1020^{\dagger}$s&  $850^{\dagger}$s &  $620^{\dagger}$s & 170$\times$20s \\
   06 &  5$\times$800s  &  5$\times$800s  &  $840^{\dagger}$s & 173$\times$20s \\
   07 &  2$\times$900s  &  6$\times$300s  &  8$\times$200s  & 144$\times$20s \\
   08 &  $900^{\dagger}$s &  $800^{\dagger}$s &  $519^{\dagger}$s & 144$\times$20s \\
   09 &  8$\times$500s  & 14$\times$300s  & 13$\times$200s  & 144$\times$20s \\
   10 &  2$\times$500s  &  6$\times$300s  &  $534^{\dagger}$s & 221$\times$20s \\
   11 &  $900^{\dagger}$s &  $800^{\dagger}$s  &  $519^{\dagger}$s & 321$\times$20s \\
   12 &  $1020^{\dagger}$s&  $800^{\dagger}$s  &  $604^{\dagger}$s & 162$\times$20s \\
   13 &  $900^{\dagger}$s &  $800^{\dagger}$s  &  $519^{\dagger}$s & 191$\times$20s \\
   14 &  $1239^{\dagger}$s&  $1098^{\dagger}$s & $2274^{\dagger}$s & 144$\times$20s \\
\hline                           
\end{tabular}
\end{table}

\begin{table}
\caption{Filter list of the LARS data. All data were obtained with NOT MOSCA, except $Ks$ (CFHT/WIRCAM), LARS06 $U$-band (NOT/ALFOSC), and the {\it HST}/WFC3 data. The ground-based data uses $U$ $\# 114$ (according to the NOT numbering scheme) for LARS01-07 and $\# 109$ for LARS 09-10. $\# 114$ has $\lambda_{eff}=347$ nm and FWHM $=38$ nm, while $\# 109$ has $\lambda_{eff}= 353$ nm and FWHM $=55$ nm. The B filter is a custom-made private filter, assigned a NOT id of $\#135$, with $\lambda_{eff}= 458$ nm and FWHM $=66$ nm for LARS 01-10. The custom-made $B$-band filter was designed to avoid the strongest emission lines. The NOT I $\# 108$ filter has $\lambda_{eff}=817$ nm and FWHM $=163$ nm. All ``F''-filter names are {\it HST}/WFC3 filters (see \citet{Ostlin2014} for details).}
\label{tab:filters}
\centering
\begin{tabular}{c c c c c }
\hline\hline                
LARS &  $U$  & $B$ & $I$ & $K$ \\   
\hline 
01 & 114  & 135  & 108   & $Ks$\\
02 & 114  & 135  & 108   & $Ks$\\
03 &      & 135  &       & $Ks$\\
04 & 114  & 135  & 108   & $Ks$\\
05 & F336W& F438W& F775W & $Ks$\\
06 & 114  & 135  & F775W & $Ks$\\
07 & 114  & 135  & 108   & $Ks$\\
08 & F336W& F438W& F775W & $Ks$\\
09 & 109  & 135  & 108   & $Ks$\\
10 & 109  & 135  & F775W & $Ks$\\
11 & F336W& F438W& F775W & $Ks$\\
12 & F336W& F438W& F775W & $Ks$\\
13 & F390W& F475W& F850LP& $Ks$\\
14 & F390W& F475W& F850LP& $Ks$\\
\hline                            
\end{tabular}
\end{table}

\section{Observations}\protect\label{sec:data}
\noindent Ground-based deep imaging data were obtained with the Nordic Optical Telescope (NOT) using MOSCA\footnote{MOSaic CAmera} on February 19-23, 2012, and February 7-8, 2013. MOSCA has a uniformly good point spread function (PSF) over its entire field of view of $7.7$ arcmin${^2}$, and a very high throughput, especially in the $U$-band, making it an efficient instrument for deep surface photometry. Additional observations using NOT/ALFOSC\footnote{Andalucia Faint Object Spectrograph and Camera} were made of LARS 06 on August 14, 17 and 18, 2012. All NIR data were obtained with the CFHT/WIRCam\footnote{Wide-field InfraRed Camera}, an imager with a field of view of $20$ square arcmin. The observations were performed using filters avoiding the strongest nebular emission lines. For the $B$-band we use our own custom made filter, $\#135$, with the transmission curve shown in the appendix (Figure \ref{fig:filter}). Tables~\ref{tab:obslog} and~\ref{tab:filters} give a more detailed description of the exposure times and filters used. {\it HST}/WFC3 imaging was used to provide similar filter coverage for all galaxies in the LARS sample for reasons explained in Section \ref{sec:calibration}.

\subsection{Data reduction}
\noindent The optical reduction was performed following the procedure in~\citet{Micheva2013a}. After the standard reduction steps of bias-subtraction and flatfielding, a pair subtraction algorithm, usually only used in NIR reductions, was applied to all data in both the optical and NIR before the average stacking of the data \citep[e.g.,][]{Melnick1999,Micheva2013a}. This technique is quite successful at flattening and minimizing the sky residuals. An additional final sky subtraction of a fitted plane with a first-order polynomial was applied on the stacked images before calibration. The optical pair subtraction and sky subtraction techniques are also described at length in~\citet{Micheva2013a}. The WIRCam $Ks$ data were reduced with the IDL Interpretor of WIRCam Images ($^\prime$I$^\prime$iwi) processing pipeline. All reduced data were resampled to a common pixel scale of $0.217$ arcsec.
\subsection{Data calibration}\protect\label{sec:calibration}
\noindent Spectrophotometric standard stars were observed and used to obtain zero points for each filter in the ground-based data. To check the viability of our calibration, we compared stellar photometry to values from the 2MASS and SDSS databases, and LARS photometry to the available {\it HST} data in \citet{Ostlin2014}. The $Ks$ photometry was found to be fully consistent with measurements from 2MASS stars in the field of view. \\

\noindent Some ground-based data were of inferior quality, likely taken during unphotometric conditions. Since these data were not salvageable, we have substituted them with {\it HST} imaging. The photometry of the remaining ground-based data is consistent with {\it HST} photometry. For LARS06 and LARS10, no ground-based $I$-band observations exist, and we complemented with {\it HST} data.

\begin{figure*}
  \begin{center}
    \makebox[\textwidth]{\includegraphics[width=16cm]{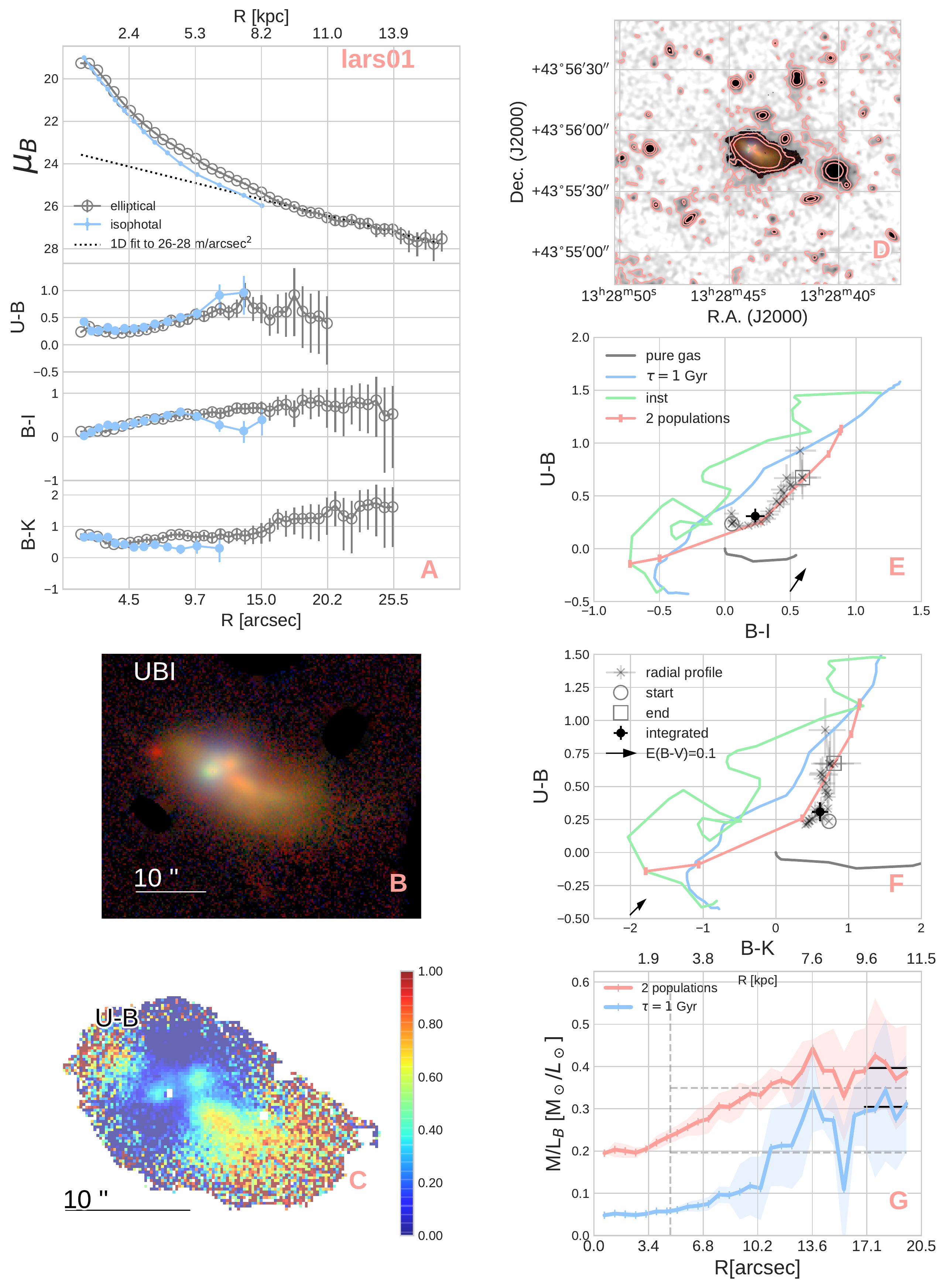}}
    \caption{LARS01: A) Surface brightness and color profiles with elliptical and isophotal integration. Where possible, a one-dimensional (1D) fit to the region $\mu_B=26-28$mag$/$arcsec${}^{-2}$ is shown (dotted line). B) RGB image with (red, green, blue) = ($I,\ B,\ U$) bands. C) $U-B$ color map. D) Deep contours with varying isophotal levels and showing the center of integration (cross).  E) and F) Color-color diagrams showing the radial color profiles (gray crosses), the total integrated color (filled black circle), instantaneous burst model (green solid line), a model with decaying SFR of e-folding time $1$ Gyr (blue solid line), a mixture of two stellar populations (red solid line), pure gas emission from an instantaneous burst (gray solid line) and a dust attenuation arrow for $E(B-V)=0.1$. G) Yggdrasil model mass-to-light radial profiles from Sect.~\ref{sec:ml}. Also indicated are the FUV $r_{cut}$ radius (vertical dashed line) and the M/L used to estimate the stellar mass of the outskirts (circle if no radial overlap, solid black line if radial overlap, gray dashed line for average).}\protect\label{fig:lars01}
  \end{center}
\end{figure*}

\begin{table*}
\caption{Integrated properties of the LARS sample. Photometry has been corrected for Galactic extinction and surface brightness dimming. Uncertainties of $\leq0.1^{\textrm{m}}$ are omitted. The central surface brightness $\mu_B$ in mag/arcsec${}^2$ and the scale length $h_R$ in kpc are also given for the 1D fit described in Sect.~\ref{sec:decompose}, as well as the position angle in degrees and the ellipticity of the elliptical radial profiles in Sect.~\ref{sec:sb}. The last column is the $3\sigma$ limiting surface brightness in the $B$-band, calculated for a $10\times10$ arcsec$^2$ area.}
\label{tab:phot}
\centering
\begin{tabular}{c l l l l l l l r l l}
\hline\hline                
LARS &  $B$  & $U-B$ & $B-I$ & $B-K$ & kpc$/\prime\prime$ & $\mu_B$ & $h_R$ [kpc] & P.A. & e & $\mu^{lim}_B(3\sigma)$\\   
\hline                       
01 & $ 15.77         $ & $ 0.33         $ & $ 0.31         $ & $ 0.67          $ & $ 0.56$ & $23.48$ & $ 4.16$  &  $-16.1$ &  $0.29$ & $28.1$ \\
02 & $ 16.77         $ & $ 0.66         $ & $ 0.45         $ & $ 0.25          $ & $ 0.60$ & $23.96$ & $ 5.97$  &  $85.6$  &  $0.61$ & $28.6$ \\
03 & $ 14.32 \pm 0.10$ &                  &                  & $ 1.76  \pm 0.10$ &         &         &          &  $32.5$  &  $0.71$ & $27.6$ \\ 
04 & $ 15.47         $ & $ 0.70         $ & $ 0.40         $ & $ 0.72          $ & $ 0.66$ &         &          &  $41.2$  &  $0.36$ & $28.2$ \\
05 & $ 16.56 \pm 0.11$ & $ 0.29 \pm 0.12$ & $ 0.22 \pm 0.16$ & $ 0.11  \pm 0.21$ & $ 0.68$ &         &          &  $-39.2$ &  $0.29$ & $26.5$ \\
06 & $ 17.49         $ & $ 0.47         $ & $ 0.35         $ & $ 0.54          $ & $ 0.68$ &         &          &  $-77.5$ &  $0.45$ & $26.1$ \\
07 & $ 16.55         $ & $ 0.52         $ & $ 0.28         $ & $ 0.61          $ & $ 0.75$ & $26.18$ & $24.03$  &  $-51.9$ &  $0.45$ & $27.7$ \\
08 & $ 14.50         $ & $ 0.90         $ & $ 0.81         $ & $ 1.85          $ & $ 0.75$ &         &          &  $ 0.0$  &  $0.00$ & $26.1$ \\
09 & $ 15.23         $ & $ 0.39         $ & $ 0.50         $ & $ 1.23          $ & $ 0.92$ & $22.39$ & $ 6.76$  &  $-84.8$ &  $0.50$ & $27.0$ \\
10 & $ 16.63         $ & $ 0.89         $ & $ 0.75         $ & $ 1.07          $ & $ 1.10$ & $21.89$ & $ 4.19$  &  $31.6$  &  $0.40$ & $27.9$ \\
11 & $ 15.56         $ & $ 0.72         $ & $ 0.62         $ & $ 1.60          $ & $ 1.58$ &         &          &  $42.1$  &  $0.76$ & $26.4$ \\
12 & $ 17.40         $ & $ 0.32         $ & $ 0.36         $ & $ 0.52          $ & $ 1.88$ &         &          &  $28.8$  &  $0.12$ & $26.0$ \\
13 & $ 16.91         $ & $-0.29 \pm 0.10$ & $ 0.05         $ & $ 1.09  \pm 0.10$ & $ 2.57$ & $21.13$ & $ 4.54$  &  $59.4$  &  $0.29$ & $26.8$ \\
14 & $ 18.83         $ & $ 0.12         $ & $ 0.02 \pm 0.14$ & $ 0.27          $ & $ 3.05$ & $21.73$ & $ 2.83$  &  $ 0.0$  &  $0.00$ & $27.0$ \\
\hline                           
\end{tabular}
\end{table*}

\section{Analysis tools}\protect\label{sec:tools}
\noindent Below we briefly describe all analysis tools we have used in order to characterize the LARS sample. Figure~\ref{fig:lars01} shows the results for LARS01. The results for the remaining LARS galaxies are in figures \ref{fig:lars02} through ~\ref{fig:lars14} in the appendix.
\subsection{Surface brightness and color profiles}\protect\label{sec:sb}
\noindent The elliptical surface brightnesses and color profiles, presented for each galaxy in panel A of Figs.~\ref{fig:lars01}, and \ref{fig:lars02} through ~\ref{fig:lars14}, were obtained using a method similar to that described at length in~\citet{Micheva2013a}. Briefly, we choose the $B$-band as a reference frame because it is the deepest, select elliptical integration parameters from the IRAF ELLIPSE fit to the faintest well-defined isophote, and obtain a radial profile with the ellipse parameters kept constant at each integration step. The same parameters are applied to all other filters, ensuring that the same physical areas are sampled at each radial step. To get a feeling for the uncertainty of the chosen integration parameters, we vary the ellipticity ($\pm0.1$) and the position angle ($\pm5\degr$) and obtain nine surface brightness profiles per filter per galaxy. These variations have the most importance for galaxies with highly elliptical shapes, of which we have only one (LARS11), and show that our chosen parameters are quite stable.\\

\noindent As a by-product of each elliptical surface brightness profile integration, we obtain the total magnitude of the galaxy by cumulatively summing the mean fluxes inside each ring. This total magnitude is presented in Table~\ref{tab:phot} for all galaxies. The uncertainty in the total magnitude we present here is then obtained from the standard deviation of the total magnitudes of the nine surface brightness profiles with varying integration parameters. We call this uncertainty $\sigma_{param}$. The composite uncertainties of each data point in a profile are a combination of the uncertainty of the mean flux level inside each elliptical ring, given by the standard deviation of the mean $\sigma_{SDOM}$, the uncertainty in the photometric zeropoint $\sigma_{ZP}$, and the uncertainty of the sky level $\sigma_{sky}$, all added in quadrature. $\sigma_{SDOM}$ is a composite of the Poisson noise and the statistical flux fluctuations across a ring. $\sigma_{sky}$ is assumed to be constant across the entire image and is obtained for each individual frame from the standard deviation of the mean values measured inside boxes placed on ``empty'' sky regions, with the size of the boxes not exceeding the area of the faintest isophote we are trying to reach. The available area for measuring the sky is defined as the inverted mask of all detected sources. Between $\sim450$ and $\sim1170$ boxes were placed on each frame to obtain $\sigma_{sky}$. This uncertainty strongly dominates the error budget at large radii. The elliptical integration surface brightness profiles were terminated when the faint end uncertainty reached $\sim1$ mag. We also verified that the PSF of the images falls off more steeply than the observed radial surface brightness profiles, i.e., that the profiles really trace the light distribution of the galaxies and not simply the shape of the PSF. The integrated photometry for all galaxies is summarized in Table~\ref{tab:phot}.\\

\noindent We present isophotal surface brightness and color profiles, which are overplotted in red in Figures~\ref{fig:lars01} through ~\ref{fig:lars14}. Our specific technique for obtaining these is also described in detail in~\citet{Micheva2013a}, but here we only note that no smoothing is involved when obtaining the faint isophotes. Similarly to the elliptical profiles, we choose the $B$-band as a reference. We obtained isophotal area masks in $0.5$ mag bins from the $B$-band and applied those masks to all other filters, once again to ensure that the physical area we sample in each filter is identical. The uncertainties in each profile data point consist of $\sigma_{ZP}$, $\sigma_{SDOM}$, and $\sigma_{sky}$ added in quadrature. $\sigma_{SDOM}$ here does not represent statistical fluctuations, since those are, by procedural design, small inside an isophote. Rather, in this case, $\sigma_{SDOM}$ accounts for the Poisson noise. The isophotal surface brightness profiles are terminated at the point where the profile starts suffering from missing pixels in ill-defined highly porous isophotes, resulting in an artificial sharp dive in radial isophotal brightness. This usually happens close to the radius at which the equivalent area of the current isophote is smaller than the area of the previous isophote. All profiles were obtained after correcting for Galactic extinction, assuming $R_V=3.1$ and the \citet{Schlafly2011} attenuation curve. Cosmological surface brightness dimming corrections have been applied.\\

\subsection{RGB images}\protect\label{sec:rgb}
\noindent Panel B of Figures~\ref{fig:lars01} and \ref{fig:lars02} through ~\ref{fig:lars14} shows the (U,B,I) tricolor images for all galaxies. We obtained these images by implementing the~\citet{Lupton2004} algorithm, with constant arcsinh stretch with $Q=3.5$ for all galaxies. This algorithm has the advantage that the color never saturates to white. Galactic extinction correction was applied to each channel before combining.

\subsection{Color maps}\protect\label{sec:colormaps}
\noindent Panel C of Figures~\ref{fig:lars01} and \ref{fig:lars02} through ~\ref{fig:lars14} shows the spatial distribution of the $U-B$ color for all galaxies. We chose to present this color map because $U-B$ more faithfully traces the young population, and a blue $U-B$ color is a sure indicator of recent star formation, especially because our $B$-band filter is designed to omit the brightest emission lines (see Appendix \ref{sec:Bfilter}). Galactic extinction correction was applied to both filters.

\subsection{Deep contours}\protect\label{sec:deepc}
Panel D of Figures~\ref{fig:lars01} and \ref{fig:lars02} through ~\ref{fig:lars14} shows the deepest contours we could obtain from these data for each galaxy. While contours up to the Petrosian radius \citep{Petrosian1976} of each galaxy are well-behaved, at fainter isophotes the increased noise prevents the contours from closing. The Petrosian radius in this case is taken to be the radius at which the Petrosian ratio $\eta=I(r)/\left<I(r)\right>$ reaches $0.2$, as defined in \citet{Bershady2000}. To bring out the faint features in the outskirts of each galaxy we smoothed the image twice by a tophat two-dimensional (2D) kernel of $5$ and $15$ pixels. These smoothed images were used to obtain a hybrid image, consisting of original unsmoothed pixels up to the Petrosian radius, then the $5$-pixel smoothed image for the region between the Petrosian radius and $25$ mag/arcsec${}^{-2}$, and then the $15$-pixel smoothed image for the region beyond $25$ mag/arcsec${}^{-2}$. The final contours, presented in panel D, were obtained from this hybrid image.

\subsection{Stellar evolutionary models}\protect\label{sec:sems}
Panels E and F of Figures~\ref{fig:lars01} and \ref{fig:lars02} through ~\ref{fig:lars14} show the $U-B$, $B-I$, and $B-K$ color predictions from the {\it{Yggdrasil}}~\citep{2011ApJ...740...13Z} stellar evolutionary model (SEM). The two single stellar population (SSP) models (blue and green solid lines) assume an exponentially declining SFR with e-folding decay rate $\tau=10^9$ yr (blue) and an instantaneous star-formation history (SFH) (green). We note that we call both of these models SSPs for simplicity, however, only the instantaneous model is truly a single stellar population model, while the model with exponentially declining SFR is rather a continuous SFR model, containing stars of various ages following the same SFH law. Both models have a gas covering fraction of $1.0$, use Padova-AGB SB99 tracks with no rotation, assume the same metallicity of $Z=0.004$ for both the gas and the stars, and a total gas mass of $10^{10}M_\odot$ available for star-formation under the assumed SFH. The exceptions are LARS08-11 and LARS13, for which a model metallicity of $Z=0.020$ was adopted to accommodate their higher observed metallicity \citep{Ostlin2014}. The model predictions were obtained using the transmission of each individual filter and the redshift for each galaxy.\\

\noindent To account for the possibility that a single SFH law may not be a good representation of the LARS galaxies, we tested mixing two SSPs with different properties (SFH, age, metallicity). There are many ways to mix stellar populations. We chose here to keep the age of both populations constant and simply vary the mass fraction of the young population. The mixture in best agreement with the data (red solid line in panels E and F) is therefore not an evolutionary track. For most LARS galaxies a mixture of a young population from the instantaneous burst model at age $5$ Myr, and an old population of $3$ Gyr old, with constant SFR for $10^8$ yr, are in good agreement with the data. For LARS01, 05, and 12, a better agreement with data is obtained if the $3$ Gyr-old population in the mixture has an exponentially declining SFR model with e-folding decay rate $\tau=10^9$ yr. The metallicity for both populations is taken to be $Z=0.004$ for the gas and stars, except for LARS08-11 and 13, where it is $Z=0.020$. The resulting red solid line connects the positions of nine mass fractions of the young population, between $1$ and $10^{-8}$ in steps of $-1$ dex.

\noindent For most galaxies the data points are consistent with the mixed population line, or lie between the mixed population and the SSP tracks. However, some points deviate, usually the points from the galaxy centers. For this reason we also present a pure gas emission track from the instantaneous burst model (gray solid line), unless otherwise explicitly stated in Section \ref{sec:notes}. This gas emission track was obtained by subtracting a model with pure stellar emission from a model with stellar and nebular emission, and represents gas emission due to an ionizing stellar population aging from $1$ to $\sim20$ Myr. This track is located in the right region for all galaxies, in the sense that it brackets the deviating points, suggesting that these regions are perhaps dominated by nebular emission.\\

\noindent These panels also show the integrated total color of each galaxy (black circle with error bars), as well as the radial color profiles (gray crosses). The dust attenuation arrow with $E(B-V)=0.1$ is for the~\citet{Cardelli1989} extinction law.

\subsection{Decomposition into young and old populations}\protect\label{sec:decompose}
\noindent In order to obtain an estimate for the strength of the starburst in each of the LARS galaxies we choose the $B$-band as the most suitable filter. We attempted to obtain a first-order approximation of the underlying old stellar population by fitting a 2D exponential disk model with GALFIT~\citep{Peng2010}, using only the $\mu_B\ge24.0$ mag/arcsec${}^{2}$ outer regions of each object. However, the majority of the LARS galaxies do not lend themselves well to an exponential disk fit. For no galaxy could we obtain a reasonable model disk which does not oversubtract parts of the inner galaxy region. Constraining the GALFIT input parameters to ensure the inner regions are not oversubtracted resulted in a disk too faint to account for the observed light in the outer regions. \\

\noindent Instead, we chose to decompose the stellar populations of the LARS galaxies using two alternative techniques. In one, we simply fit an exponential disk to the 1D radial surface brightness profile, following the exact procedure as for the $45$ BCGs in~\citet{Micheva2013a,Micheva2013b}. Similar to the BCGs, we fit the range $26<\mu_B<28$ mag/arcsec${}^{2}$ with an exponential disk profile. This is not possible for every LARS galaxy, since the constraint is that a line fit should not be brighter than the observed profile at any point. Where possible, the 1D exponential disk is given with a dotted line in panel A of Figures~\ref{fig:lars01} and \ref{fig:lars02} through ~\ref{fig:lars14}. We note that it is difficult to determine the physical meaning of a scale length obtained in this way, as the profile is clearly not an exponential disk. This scale length is only a representation of how quickly the average flux decreases in the outer regions, and allows for a meaningful comparison to the BCGs 1D fit results, which were obtained with the same technique. The central surface brightness and scale length of the 1D fits are summarized in Table~\ref{tab:phot}.\\

\noindent The second technique involves using a blue filter dominated by light from young stars to define the 2D region occupied by the young population. We call this a ``mountain top'' technique because it clips the frustum of the 2D flux distribution. To obtain the extent of this region, we use FUV {\it HST} data, which samples only the young population, to find the radius at which the FUV isophotal surface brightness drops to a corresponding SFR surface density (SFRD) of $0.01$ M$_\odot$/yr/kpc$^2$. This SFRD is approximately where the slope of the Kennicutt-Schmidt law changes from efficient to inefficient star formation, and is in line with what is obtained from recent resolved studies of actively star-forming galaxies \citep[e.g., the Whirlpool galaxy, M51;][]{Bigiel2008}. The limit on the SFRD assumes zero extinction which is a reasonable approximation at large radii. The FUV {\it HST} data are in filter $F140LP$ for LARS01-12, and $F150LP$ for LARS13-14 \citep{Ostlin2014}. Using the SFR-FUV flux calibration of \citet{Kennicutt1998}, $SFR(UV)[M_\odot/yr]=1.4\times10^{-28}L_\nu$[erg/s/Hz], this gives a cutoff isophote of $\mu_{FUV}=23.53$ mag/arcsec$^2$. The area brighter than $\mu_{FUV}$ gives the equivalent cutoff radius $r_{cut}=\sqrt{\textrm{area}/\pi}$, which in turn provides the $B$-band isophote at the base of the mountaintop, $\mu_B^{cut}$. Isophotes fainter than $\mu_B^{cut}$ are discarded, thus removing all contribution from outside of the mountain top. To remove the contribution of the old population underneath the mountain top, we assume a constant flux equal to the first isophote outside of the mountain top, and subtract it from every pixel in the mountain top region. The $B$-band burst fraction $\zeta_{B}^{FUV}$, where the superscript 'FUV' indicates that the cutoff radius is based on FUV data, is then the ratio of the mountain top region to the total flux of the galaxy, namely $\zeta_{B}^{FUV}=\sum{F(r\leq r_{cut})}/F_{tot}$. The obtained $B$-band burst fractions and cutoff radii are summarized in Table~\ref{tab:mountaintop}. The table also shows the fraction of FUV flux, $\zeta_{FUV}^{FUV}$, and H$\alpha$ flux, $\zeta_{H\alpha}^{FUV}$, inside of the mountaintop region. The cutoff radius is indicated for each galaxy with a dashed vertical line in panel E of Figures~\ref{fig:lars01}, and \ref{fig:lars02} through ~\ref{fig:lars14}.\\

\noindent In order to compare to the BCGs, where FUV data are not largely available, we perform the same mountain top technique but using the isophotal $U$-band radial flux profile to define the cutoff radius at the surface brightness corresponding to SFRD$=0.01$ M$_\odot$/yr/kpc$^2$. Here we have assumed a flat spectrum in $f_\nu$, so that $L_\nu^{FUV}=L_\nu^{U}$. The FUV and $U$-band isophotal profiles are relatively similar, and give very similar cutoff radii, as demonstrated in Table~\ref{tab:mountaintop}. The $B$-band burst fractions, in this case denoted $\zeta_{B}^U$ to indicate that the cutoff is obtained from $U$-band, are also shown in Table~\ref{tab:mountaintop}. To enable the comparison, we repeat this procedure for all BCGs with existing $U$-band data in~\citet{Micheva2013a,Micheva2013b}. 

\subsection{M/L radial profiles}\protect\label{sec:ml}
Panel G of Figures~\ref{fig:lars01} and \ref{fig:lars02} through ~\ref{fig:lars14} shows mass-to-light ($M/L_B$) radial profiles for each galaxy. For each radial color data point in panel E we computed the minimal distance to the Yggdrasil tracks, and assigned it the corresponding model M/L$_B$ ratio. For all galaxies we obtain one M/L profile for an SSP model (blue line) and one for the two-population mixture (red line). These are the same models shown in panel E and described in Section \ref{sec:sems}. The error bars at each radial step are the standard deviations of all model M/L ratios consistent with a given color point within its error bars. This means that as the uncertainties of the color data points increase towards the outskirts, so does the uncertainty on which modeled M/L ratio is closest to the data. \\

\noindent Using these M/L profiles we can estimate the stellar mass in the outskirts of the LARS galaxies. The NOT $B$-band reaches $\sim28$ mag arcsec$^{-2}$, allowing for an estimate of the mass within the $26\mbox{-}28$ mag arcsec$^{-2}$ region. The radial M/L profiles are obtained from a $U-B$ vs. $B-I$ diagram (panel E), and hence the maximum radial length one can reach is determined by the common length of all three filters, which never reaches $28$ mag arcsec$^{-2}$. We therefore assume that at larger radii, corresponding to $26\mbox{-}28$ mag arcsec$^{-2}$, the M/L is constant and equal to the last profile data point (black circle in panel G). In the case of partial overlap between the $B$-band surface brightness and the M/L radial profiles in the region of interest we take the average M/L of the overlap as representative of the ratio of the outskirts (black solid horizontal line in panels G). The resulting $L_B$ luminosities, M/L ratios, and masses are listed in Table \ref{tab:mass}. While we show M/L profiles with both the SSP and two population mixture in panel G, in the table we only list the M/L obtained from the model closest to the observations.

\noindent Using our mountaintop method to separate the burst from the rest of the galaxy, we can also estimate the stellar mass that lies outside of the mountaintop. For consistency, we use a common surface brightness isophote of $25$ mag arcsec$^{-2}$ for all galaxies, and estimate the average M/L ratio (gray horizontal dashed line in panel G) in the region $\mu_B^{cut}\leq\mu_B\leq25$. The mass estimates and model M/L ratios are summarized in Table \ref{tab:mass}.

\begin{table*}
  \begin{minipage}{186mm}
    \caption{Burst fractions $\zeta$ with the mountain top method, using the FUV and $U$-band data to define the radius $r_{cut}$ at which SFRD$=0.01$M$_\odot$/yr/kpc$^2$. $\zeta_B$, $\zeta_{FUV}$,  $\zeta_{H\alpha}$, and $\zeta_U$, are the burst fractions in $B$-band, FUV, H$\alpha$, and $U$-band, respectively. The corresponding cut-off surface brightness, $\mu_B^{cut}$, is given for reference. }
    \protect\label{tab:mountaintop}
    \centering
    \begin{tabular}{@{}cccccc|cccc@{}}
      & \multicolumn{5}{c}{from FUV} &  \multicolumn{4}{c}{from $U$-band}\\
      LARS & $r_{cut}$ & $\mu_B^{cut}$ & $\zeta_B^{FUV}$ &$\zeta_{FUV}^{FUV}$&$\zeta_{H\alpha}^{FUV}$& $r_{cut}$&$\mu_B^{cut}$& $\zeta_B^U$ & $\zeta_U^U$  \\
           & kpc    & mag arcsec${}^{-2}$&               &                 &                & kpc      &   mag arcsec${}^{-2}$&      &           \\   \hline\hline
01   &  $2.69$        &   $22.06$   &  $0.55$         &  $0.88$        &  $0.94$           & $3.51$        &  $22.90$  & $0.69$      & $0.75$  \\    
02   &  $2.00$        &   $22.34$   &  $0.32$         &  $0.73$        &  $0.84$           & $2.33$        &  $22.63$  & $0.37$      & $0.57$  \\    
03   &  $2.34$        &   $20.42$   &  $0.09$         &  $0.44$        &  $0.85$           & \ldots        &  \ldots   & \ldots      & \ldots   \\  
04   &  $4.21$        &   $22.08$   &  $0.40$         &  $0.76$        &  $0.95$           & $4.69$        &  $22.40$  & $0.48$      & $0.66$    \\  
05   &  $2.20$        &   $23.05$   &  $0.91$         &  $0.91$        &  $0.93$           & $2.14$        &  $22.95$  & $0.90$      & $0.92$  \\    
06   &  $2.05$        &   $22.52$   &  $0.16$         &  $0.49$        &  $0.72$           & $2.36$        &  $22.69$  & $0.23$      & $0.34$  \\    
07   &  $2.33$        &   $21.79$   &  $0.54$         &  $0.87$        &  $0.95$           & $2.82$        &  $22.50$  & $0.66$      & $0.78$   \\   
08   &  $5.10$        &   $21.66$   &  $0.57$         &  $0.70$        &  $0.87$           & $6.19$        &  $22.17$  & $0.66$      & $0.75$   \\   
09   &  $6.70$        &   $22.12$   &  $0.55$         &  $0.80$        &  $0.94$           & $7.38$        &  $22.60$  & $0.66$      & $0.80$   \\   
10   &  $3.12$        &   $21.56$   &  $0.35$         &  $0.66$        &  $0.82$           & $4.24$        &  $22.33$  & $0.51$      & $0.61$   \\   
11   &  $8.40$        &   $22.06$   &  $0.63$         &  $0.78$        &  $0.91$           & $9.68$        &  $22.71$  & $0.76$      & $0.78$   \\   
12   &  $4.44$        &   $23.32$   &  $0.95$         &  $0.91$        &  $0.96$           & $4.07$        &  $22.86$  & $0.92$      & $0.90$   \\   
13   &  $7.40$        &   $22.35$   &  $0.63$         &  $0.80$        &  $0.89$           & $8.49$        &  $22.89$  & $0.74$      & $0.77$   \\   
14   &  $5.03$        &   $23.25$   &  $0.89$         &  $0.90$        &  $0.92$           & $5.24$        &  $23.41$  & $0.90$      & $0.88$   \\   
    \end{tabular}
  \end{minipage}
\end{table*}

\begin{table*}
  \begin{minipage}{186mm}
    \caption{Stellar mass and mass-to-light ratio estimates for two regions: from the base of the mountaintop to $\mu_B=25$ mag arcsec${^{-2}}$, and $26\leq \mu_B\leq 28$ mag arcsec$^{-2}$. The model M/L$_B$ ratios are averages across the region for the former case, and either averages or extrapolation for the latter (see text). M/L$_B$ ratios from the two-population mix are marked with $\dagger$. Columns $M_\star/M_\star^{SED}$ represent the fraction that the estimated mass makes up of the total stellar SED mass modeled by \citet{Hayes2014}. }
    \protect\label{tab:mass}
    \centering
    \begin{tabular}{@{}lcccc|cccc|c@{}}
      & \multicolumn{4}{c}{$\mu_B^{\textrm{cut}}\leq \mu_B\leq 25$} &  \multicolumn{4}{c}{$26\leq \mu_B\leq 28$} &\\
      LARS & $\left<\textrm{M/L}\right>$ & $L_B$  &  $M_\star$ & $\frac{M_\star}{M_\star^{SED}}$ &   M/L$_B$ & $L_B$ & $M_\star$ &  $\frac{M_\star}{M_\star^{SED}}$  & $M_\star^{SED}$\\
      &$M_\odot/L_\odot$                                & $10^9L_\odot$              &  $10^9M_\odot$ &   & $M_\odot/L_\odot$  & $10^9L_\odot$ &  $10^9M_\odot$   & &$10^9L_\odot$   \\\hline\hline
      $01^\dagger$ & $ 0.35$ & $ 3.48$ & $ 1.22$ & $ 0.20$ & $ 0.40$ & $ 0.57$ & $ 0.23$ & $ 0.04$ & $6.10$\\
      $02$ & $ 0.30$ & $ 3.22$ & $ 0.97$ & $ 0.41$ & $ 0.65$ & $ 0.62$ & $ 0.40$ & $ 0.17$& $2.35$\\
      $03$ & \ldots  & \ldots  &  \ldots & \ldots  &  \ldots & \ldots  & \ldots  & \ldots & \ldots\\
      $04$ & $ 0.30$ & $ 9.82$ & $ 2.91$ & $ 0.23$ & $ 0.28$ & $ 0.52$ & $ 0.15$ & $ 0.01$ & $12.90$\\
      $05^\dagger$ & $ 0.28$ & $ 1.76$ & $ 0.49$ & $ 0.11$ & \ldots  & $ 0.34$ & \ldots  & \ldots & $4.27$\\
      $06$ & $ 0.26$ & $ 2.04$ & $ 0.54$ & $ 0.26$ & \ldots  & $ 1.20$ & \ldots  & \ldots & $2.09$\\
      $07$ & $ 0.22$ & $ 3.99$ & $ 0.87$ & $ 0.18$ & $ 0.19$ & $ 0.79$ & $ 0.15$ & $ 0.03$ & $4.75$\\
      $08$ & $ 1.30$ & $19.64$ & $25.52$ & $ 0.27$ & \ldots  & $ 4.04$ & \ldots  & \ldots & $93.30$\\
      $09$ & $ 0.66$ & $28.32$ & $18.78$ & $ 0.37$ & $ 0.77$ & $ 1.95$ & $ 1.50$ & $ 0.03$ & $51.00$\\
      $10^\dagger$ & $ 1.19$ & $10.51$ & $12.55$ & $ 0.58$ & $ 1.22$ & $ 1.38$ & $ 1.68$ & $ 0.08$ & $21.50$\\
      $11$ & $ 0.51$ & $92.40$ & $47.18$ & $ 0.39$ & \ldots  & $ 8.41$ & \ldots  & \ldots & $121.00$\\
      $12$ & $ 0.14$ & $ 7.18$ & $ 1.02$ & $ 0.14$ & \ldots  & $ 0.98$ & \ldots  & \ldots & $7.41$\\
      $13^\dagger$ & $ 0.62$ & $34.87$ & $21.62$ & $ 0.37$ & \ldots  & $15.37$ & \ldots  & \ldots & $59.20$\\
      $14^\dagger$ & $ 0.17$ & $ 6.55$ & $ 1.12$ & $ 0.64$ & \ldots  & $ 3.40$ & \ldots  & \ldots & $1.75$\\
    \end{tabular}
  \end{minipage}
\end{table*}

\begin{figure*}
  \begin{center}
    \makebox[\textwidth]{\includegraphics[width=15.6cm]{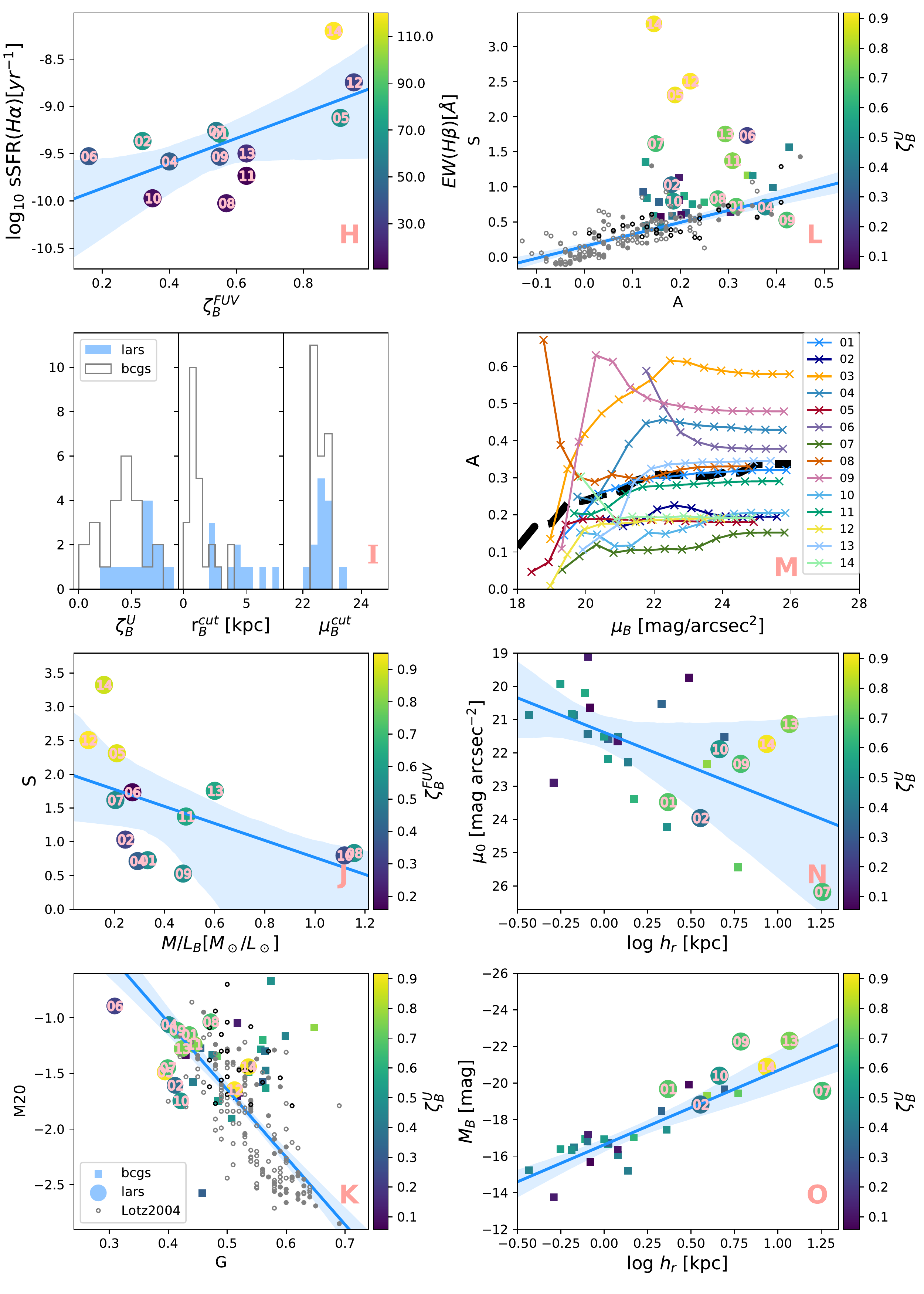}}
    \caption{
      H) LARS sSFR(H$\alpha$) and burst fraction $\zeta_B^{FUV}$, color-coded by EW($H\beta$).
      I) comparison between LARS and BCG burst fractions, cutoff radii [kpc] and $B$-band cutoff surface brightness.
      J) LARS clumpiness S vs. median M/L$_B$ ratio, color-coded by $\zeta_B^{FUV}$.
      K) M20 vs. Gini coefficients, color-coded by $\zeta_B^{U}$. The \citet{Lotz2004} sample contains E/S0-Sd (gray open circles) from \citet{Frei1996}, SDSS $u$-band selected galaxies (gray filled circles) and dIrr (open black circles) from \citet{vanZee2001}.
      L) As in panel K but for clumpiness S vs. asymmetry A. 
      M) Asymmetry A as a function of $B$-band isophotal surface brightness $\mu_B$. The average from all galaxies is indicated with a dashed black line.
      N) Central surface brightness $\mu_0$ and scale length [kpc] from the 1D fit (Sect. \ref{sec:decompose}).
      O) Absolute $B$-band magnitude and 1D fit scale length.
      All panels: The shaded area is the confidence interval of the linear regression (blue solid line). BCG photometry has been converted from Vega mag to AB mag. Galactic extinction and surface brightness dimming has been applied. Numbers inside markers indicate the LARS number.}\protect\label{fig:discuss}
  \end{center}
\end{figure*}
\section{Discussion}\protect\label{sec:discuss}
The morphological diagnostics we have examined reveal two main morphologies in the LARS sample. One is seemingly compact with irregular outer isophotes (LARS12, LARS13, and LARS14), while the other type is characterized by an extended disturbed appearance (all remaining LARS galaxies). The compact morphology galaxies are notably at the highest redshifts in the sample, however, down to the comparable isophotal levels, their morphologies are intrinsically different from the lower redshift LARS. This is likely a selection effect, since all three of these galaxies were selected from a sample of compact sources with high far-UV luminosity and surface brightness \citep{Ostlin2014}.\\

\noindent The ground-based data reach on average $\sim1.5$ magnitudes fainter than the {\it HST} data, as seen by the radial profiles and the $3\sigma$ limiting surface brightness in Table \ref{tab:phot}. With NOT $B$-band we can probe the $26\mbox{-}28$ mag arcsec$^{-2}$ region of LARS01, 02, 04, 07, 09, and 10. These regions contain substantial stellar mass, listed in Table \ref{tab:mass}. For most of these galaxies, these mass estimates amount to only a few per cent of the total stellar masses in \citet{Hayes2014}. For LARS02 and 10, they amount to $17$ and $8\%$, respectively, suggesting that their total mass estimates, which do not include the $26\mbox{-}28$ mag arcsec$^{-2}$ region, are underestimated by that amount. \\

\noindent Morphologically, the star formation appears to be merger-driven for most galaxies in the sample. LARS01 and LARS10 have peculiar morphologies, with asymmetric outer isophotes. LARS12 and 14 are compact but display similarly irregular outer isophotes. \citet{Pardy2014} identify LARS03, 04, 06, 09, and 11 as part of interacting systems and the same is likely for LARS13, with its dual-core morphology. LARS07, 08, and 10 also show telltale post-merger features, such as shell-like structures and tails. Merger-driven star formation is consistent with our observations. In the LARS galaxies, young blue regions are often found in areas where significant star formation would not normally take place were the galaxy in isolation, namely scattered in the outer regions. This is seen in the $U-B$ color plots in panel C of Figs.~\ref{fig:lars01} and \ref{fig:lars02} through~\ref{fig:lars14}. This could be due to the precursors to the observed mergers having comparable mass \citep{Cox2008}. The irregular morphology of the scattered star forming regions also suggest that the LARS galaxies are in the early stages of the merger event since as mergers approach coalescence, the star formation tends to become nuclear \citep{Powell2013}. \\

\noindent In general, the isophotal surface brightness profile by design has the steepest possible slope, while the slope of the elliptical profile is likely to be similar or flatter, depending on how well the elliptical rings trace the shape of the isophotes. Some degree of disagreement between the two types of profiles is therefore to be expected. For the LARS galaxies the isophotal and elliptical $U-B$ and $B-I$ color radial profiles are in good agreement with each other, indicating that the isophotes are similarly shaped across the three colors. This is not surprising since all three bands experience significant contributions from the same (young) population. If the isophotal and elliptical $B$-band surface brightness profiles match fairly well, the comparison between an isophotal and elliptical color profile involving the $B$-band and a complementary filter will tell us if the spatial distribution of the population dominant in the complementary band is the same as in $B$. Specifically, the $B-K$ color profiles are more sensitive to the presence of an old population. For most LARS galaxies the isophotal $B-K$ color profile closely traces the slope of the elliptical profile, suggesting that the central isophotes of both the young and old populations have elliptical-like shapes. The isophotal and elliptical $B-K$ LARS03 profiles in Fig. \ref{fig:lars03} are not a good contrary example of divergent optical and NIR morphologies because no isophote at any radial distance in the composite structure of LARS03 is well approximated by a single central ellipse either in the $B$ or in the $K$-bands.\\

\noindent For many LARS galaxies, the SSP models of any age and metallicity are inconsistent with the central bright isophotes, where the current starburst typically resides. In some cases the inconsistency extends to the entire radial color profile. For these galaxies, the two-population mixture of a young ($5$ Myr) and an old ($3$ Gyr) population, with young mass fractions typically in the range $0.001$-$0.01$, is able to reconsile the observations to model predictions. The outskirts of LARS02, 04, 06, 07, 08, and 12 are well matched by an old SSP of $1\mbox{-}2$ Gyrs, except LARS12, for which the SSP is younger, in the range $0.5\mbox{-}1$ Gyr. For LARS01, 05, 09, 13, and 14, the integrated galaxy color is consistent with the mixed-population models, while for LARS02, 04, and 12 the integrated colors are well-matched by SSP models. The rest of the LARS sample has integrated colors in between the SSP and two-populations models, but can be reconciled with the latter if dust correction is applied. The required dust attenuation suggested by the models is $E(B-V)=0.1$ to $\sim0.7$, which is in agreement with previous findings from SED modeling using the {\it HST} data~\citep{Hayes2014}.\\

\noindent While we are able to explain most of the observations invoking a mixture of two populations, some central colors still appear to be deviating, most prominently LARS08 and LARS11. Together with LARS03, these are the dustiest galaxies in the sample and therefore the hardest to explain with pure tracks and without proper fitting. It is possible that another combination of stellar populations and metallicities, or another recipe for obtaining the mixture, will be able to match these points. These deviating colors could be reconciled with pure gas emission tracks, modulo a dust attenuation correction, however, at the ages of the LARS galaxies it is not realistic for a galaxy region to be completely dominated by nebular emission. \\

\noindent In Figure \ref{fig:discuss} we show diagnostics typically used to study the morphology of galaxies - the asymmetry A$_{180}$, clumpiness S, Gini and M20 coefficients, as well as the burst fraction $\zeta_B$ (Section \ref{sec:decompose}). Where a correlation is observed, a simple linear regression is plotted (solid blue line), with the size of the confidence interval for the regression estimate (shaded area) given by the standard deviation of the observations in each bin.\\

\noindent We first note that the mountain top method gives a workable estimate of the burst fraction. With this crude method we obtain higher specific SFR (sSFR) for the LARS galaxies \citep{Hayes2014} with higher burst fractions (panel H). We then compare the burst fraction properties between the LARS and BCG samples (panel I), where we use $\zeta_B^U$ as FUV data is unavailable for the BCG galaxies. As we noted in Sect. \ref{sec:decompose}, however, the FUV and $U$-band isophotal surface brightness profiles are quite similar and give similar cutoff radii and $B$-band cutoff isophotes (Table \ref{tab:mountaintop}). The three histograms in panel I show the burst fraction (left), the cutoff radius (middle), and the cutoff $B$-band surface brightness in units of mag arcsec$^{-2}$. The range of $\zeta_B^U$ is similar for BCGs and LARS, with an average of $\zeta_B^U=0.65$ for LARS and $\zeta_B^U=0.40$ for BCGs. The cutoff radius is smaller for BCGs than for LARS, with averages $r_B^{cut}=1.3$ kpc and $4.9$ kpc, respectively, implying that the mountaintop region, and hence the star formation region, is more spatially extended in LARS. The cutoff $B$-band isophote is on average similar for BCGs and for LARS, with averages $\mu_B^{cut}=22.5$ and $22.7$ mag arcsec$^{-2}$, respectively. This implies that at the cutoff radius BCG and LARS galaxies are similarly blue in $U-B$, since the cutoff $U$-band surface brightness is the same for both.\\

\noindent We have tested all morphological parameters (A$_{180}$, S, G, M20), obtained from the $B$-band, for any correlation with the burst fraction, and with properties like SFR, sSFR, $f_{esc}(\textrm{Ly}\alpha)$, and equivalent width of Ly$\alpha$. There is a weak trend for the clumpiness to increase with burst fraction, which can be seen in panels J and L, where the markers are color-coded by $\zeta_B^{FUV}$. Galaxies with high burst fractions could be mergers, and in turn are likely clumpier than less strongly star-forming galaxies. Another positive correlation one intuitively expects is for the asymmetry to increase with burst fraction (or equivalently sSFR); if the starburst is due to a merger event both the burst fraction and the asymmetry should increase as a result. A weak correlation is observed in panel L, with the high $\zeta_B^{FUV}$ galaxies located on the high asymmetry end of the distribution. M20 and Gini coefficients show no discernible correlation with $\zeta_B^{FUV}$ (panel K), suggesting that the amplitude of the merger event is not the major factor governing the behavior of these morphological parameters.\\

\noindent We found no correlation between Ly$\alpha$ escape fraction or equivalent width and any of the morphological parameters of asymmetry A, clumpiness S, Gini and M20 coefficients, consistent with the results of \citet{Guaita2015}. Our morphological parameters were obtained from optical $B$ band images, dominated by stellar continuum emission. Significant differences in morphology between Ly$\alpha$ and stellar continuum (in FUV) were found by \citet{Hayes2014}. Specifically, Ly$\alpha$ surface brightness radial profiles in the LARS sample seem flatter than the FUV continuum counterpart toward the center of the galaxies, seemingly more consistent with an exponential disk with Sersic index typically $n=1\mbox{-}2$. Additionally, \citet{Hayes2013} show that in the LARS sample, the Ly$\alpha$ light is significantly more spatially extended than the stellar continuum in FUV. Given this observed difference in morphology between Ly$\alpha$ emission and stellar continuum, it is not surprising that our morphological parameters, derived from the spatial distribution of the stellar continuum emission, do not correlate with any Ly$\alpha$ properties.\\

\noindent A relation is observed in panel J, where the clumpiness S increases with decreasing M/L ratio, where the latter is taken to be the median over each galaxy. This relation is expected because galaxies with lower M/L ratios have a dominant young population, likely due to an ongoing burst. In such a scenario the burst is expected to be clumpy, which increases the value of S. \\ 

\noindent The asymmetry A we have plotted in panel L is the minimal rotational composite asymmetry, measured across the entire galaxy region in the $B$-band, and dominated by the strong flux in the flocculent starburst. As such, the asymmetries for the LARS galaxies are very similar in value to what we obtained for the sample of BCGs in~\citet{Micheva2013a,Micheva2013b}. In fact, all morphological diagnostics like asymmetry, clumpiness, Gini and M20 coefficients, show similar values for both the LARS galaxies and the BCGs. This is demonstrated in panels K and L in Fig.~\ref{fig:discuss}, which shows that both samples occupy the same regions in this parameter space. There is, notably, a separation between LARS/BCGs and a control sample from \citep{Lotz2004}. The Lotz sample contains normal elliptical galaxies (E-Sd) from \citet{Frei1996}, which are representative of standard morphology, SDSS DR1 $u$-band selected galaxies ($u<14^{\textrm{m}}$), and dwarf irregulars (dIrr) from the \citet{vanZee2001} sample. LARS/BCGs occupy the extreme ends of these distributions, for example, with high asymmetry and high clumpiness in panel L. The dIrr and some of the more extreme normal galaxies from the Lotz sample overlap with the position of the LARS and BCGs in these plots. We note further that while these morphological parameters can separate irregular star-forming galaxies from regular ones, none of them capture the nuances in morphology found among starbursting galaxies. In the region occupied by LARS/BCGs, the seeming correlation between the morphological parameters collapses and becomes a scatter plot. \\

\noindent In an effort to characterize in greater detail the morphological structure of the LARS galaxies, in panel M in Fig.~\ref{fig:discuss} we show the isophotal progression of the asymmetry for all LARS galaxies. The average behavior of the sample shows lower asymmetries in the center (brighter isophotes), and a tendency for the asymmetry to increase towards the outskirts of the galaxies (fainter isophotes). The exceptions are LARS06, LARS08, for which the {\it HST} images only cover the central part of the galaxy and not the tail, and LARS14, which is very compact and Green Pea-like. This representation captures what is immediately obvious from the deep contours in panel D of Figures ~\ref{fig:lars01}, and \ref{fig:lars02} through~\ref{fig:lars14}, namely, that the outskirts of the LARS galaxies are on average highly irregular and show no hint of an underlying extended and undisturbed population. This is contrary to what we found for the BCG sample. In both~\citet{Micheva2013a,Micheva2013b} a property common to many BCGs that we examined was the tendency for the outer regions to display a highly regular morphology, which implies that the host population of the underlying galaxies has not undergone a significant dynamical disturbance for a number of gigayears. This difference in morphology is likely to reflect differences in the history of dynamical events. In the case of the LARS galaxies, the implication is that the merger event is more recent, more major, or both. In contrast, in the average BCG the event is either an accretion, or a much more evolved major merger. \\

\noindent Other physical parameters, consistent with this distinction between LARS and BCGs, are the absolute $B$-band magnitude $M_B$, central surface brightness $\mu_0$, and scale length $h_r$ of the 1D fit, described in Sect. \ref{sec:decompose}. The 1D exponential disk fits of the LARS galaxies have fainter $\mu_0$ and larger $h_r$ (panel N), albeit with much scatter in $\mu_0$, and they are on the bright, extended end of the distribution in $M_B$-$h_r$, seamlessly continuing the trend set by the BCGs (panel O). \\

\noindent An additional argument can be made from the color radial profiles in panel A of Figs. \ref{fig:lars01} and \ref{fig:lars02} through \ref{fig:lars14}. The $U-B$ colors sample the Balmer break, and are therefore a good indicator of age. If the population is very young, the Balmer break will not be pronounced, resulting in bluer $U-B$ colors. In the low-luminosity BCG case the $U-B$ color profiles tended to redden with increased radius. In the case of LARS, eight out of $14$ galaxies show either flat profiles, or become bluer in the color with radius, indicating the presence of a dominant younger population even at higher radii. These are additional signs of the early stages of an ongoing major wet merger. To compare to BCGs, we consider that $\sim80\%$ of BCGs ($35$ out of $43$ galaxies in \citet{Micheva2013a,Micheva2013b}) have regions of intense star formation, embedded in an underlying old and extended stellar population, which has symmetric, regular outer isophotes. This underlying population is called a ``host'' population, to reflect that it is seemingly unaffected by the ongoing star formation, preserving its symmetric, regular morphology, especially at large distances from the star formation regions. In contrast, it is difficult to speak of a ``host'' population in LARS galaxies. This does not mean that an old underlying population is not present, but rather that it cannot be probed even with observations of comparable depth to the BCGs, because most of the LARS galaxies surface area shows signs of star formation and recent dynamical events. \\

\noindent We note, however, that the BCGs are a highly inhomogeneous class of galaxies, with morphologies ranging from nE and iE (nuclear or irregular central burst, respectively, with elliptical outer isophotes), to iI-C and iI-M (off-center nucleus in a cometary host or an apparent merger, respectively), as established by \citet{Loose1986}, for example. They are, on average, small, actively star forming and have low metallicities \citep[e.g.,][]{Kunth1986, Kunth2000}. The BCG class overlaps with dwarf irregulars and magellanic galaxies, but has on average more active star formation. \citet{Telles1997} further classify BCGs into two types, with type I being more luminous and with an irregular morphology, and type II being fainter and having a regular, symmetric morphology in the outskirts. Some BCGs, notably luminous BCGs like Haro11 \citep{Micheva2010}, and Eso249-31 and Tol1457-262 \citep[][both with no ground-based $U$-band]{Micheva2013a}, would classify as Type I in this nomenclature, and are morphologically indistinguishable from the LARS galaxies at the faintest levels we can reach of $26$-$28$ mag arcsec${}^{-2}$. In fact, only $27\%$ of the luminous BCGs ($M_B\leq-18$ AB) have regular outer isophotes in \citet{Micheva2013a,Micheva2013b}, while $\sim80\%$ of the faint BCGs from the same sample show regular outer isophotes. Luminous BCGs with disturbed morphologies are therefore expected to resemble LARS morphologies to various degrees.

\section{Conclusions}\protect\label{sec:conclusions}
We have investigated deep surface photometry imaging of the LARS sample in broadband $UBIKs$ bands using ground-based data with $3\sigma$ limiting surface brightness of $\mu_B\sim28$ mag arcsec$^{-2}$. For galaxies lacking ground-based data we used available {\it HST} data of shallower depth of $\mu_B\sim26.5$ mag arcsec$^{-2}$. We obtain elliptical and isophotal radial surface brightness and color profiles, isophotal asymmetry profiles, deep contour and color maps. Stellar evolutionary models are compared to both integrated and radial colors, and suggest that for many of the LARS galaxies a single stellar population model cannot explain the observed central colors. Instead, a mixture of two populations can explain the data for varying mass fractions of the young population. For the LARS galaxies, we successfully reconciled the observed colors with model predictions for a combination of a $5$ Myr-old instantaneous burst and a $3$ Gyr-old population with either an exponentially declining or constant SFR.\\

\noindent Morphological diagnostics like asymmetry A, clumpiness S, M20 and Gini coefficients can separate the LARS sample from a control sample of normal (E-Sd) galaxies but not from other irregular mergers like the BCGs or dIrrs. For the LARS sample, we found no correlation between any of these morphological parameters and the properties of Ly$\alpha$, such as escape fraction and equivalent width. One clear difference between LARS and faint BCGs of Type II in the nomenclature of \citet{Telles1997} is that the deep contours of LARS galaxies remain irregular down to the faintest isophotes we can probe, which are of comparable faintness to what is available for BCGs, and hence suggest a genuine difference between the two samples. This difference is illustrated by the isophotal behavior of the asymmetry parameter, which progresses from lower to higher asymmetry with decreasing isophotal brightness. The opposite is seen in Type II BCGs, where the presence of an underlying extended old host population with regular disk-like shape drives the asymmetry to lower values. This difference is not seen when looking at the asymmetry of the entire galaxy region; one must instead construct an isophotal asymmetry profile. Luminous BCGs of Type I are, however, on average morphologically indistinguishable from the LARS galaxies.\\

\noindent Comparison of our burst fraction estimate $\zeta_B$ for BCG and LARS galaxies suggests that the burst region occupies a larger surface area in the LARS sample than in the BCG sample. Further, no regular underlying host population can be detected for LARS in deep optical or NIR broadband imaging, contrary to what is found in typical BCGs. \\

\noindent We conclude that the LARS galaxies are in the early stages of major mergers, and in particular, merging stages earlier than what is observed in the average BCG. 

\appendix
\section{Filter \#135}
\begin{figure}
  \begin{center}
    \includegraphics[width=\hsize]{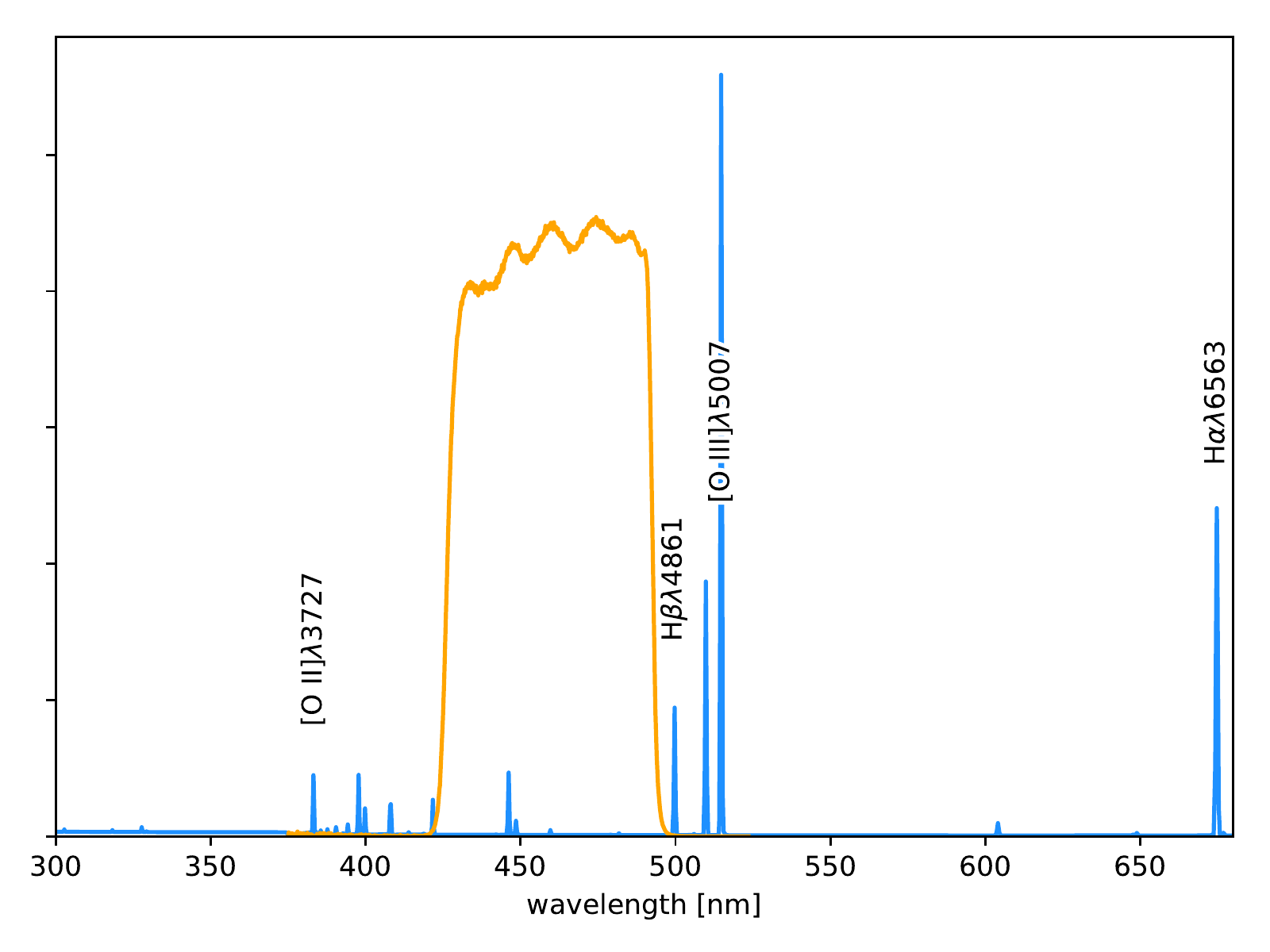}
    \caption{Filter transmission for  \#135. An example galaxy at $z=0.028$ is shown for reference.}\protect\label{fig:filter}
  \end{center}
\end{figure}

\protect\label{sec:Bfilter}
The $B$-band filter with ID $\#135$ is a custom-made filter designed to provide a clean measurement of the blue continuum redwards of the $4000$\AA-break and to avoid the strongest emission lines, such as H$\beta$ and [O III] $\lambda\lambda4959,5007$. This is achieved in the redshift range from $z=0.025$, where H$\beta$ is shifted out of the bandpass, to $z\sim0.08$, where the $4000$\AA-break shifts into the bandpass. The filter transmission is shown in Fig. \ref{fig:filter}, with an example template 3 Myr-old population at the lowest redshift of the sample of $z=0.028$ for LARS01.

\section{Notes on individual galaxies}\protect\label{sec:notes}
In this section we briefly describe some notable observations the analysis tools in Sect.~\ref{sec:tools} reveal for each galaxy.
\subsection*{LARS01}
The central region seems dominated by a young population, as indicated by the behavior of the color radial profiles. This region is extended and irregular, as suggested by the $U-B$ spatial map. The three contours in panel D of Fig. \ref{fig:lars01} are at $24.5, 25.6$, and $26.5$ mag/arcsec${}^2$, respectively. The faintest contour is highly irregular, hinting at a recent merger event. Comparison to the SEMs in panels E and F suggests that at any radial distance from the center, a SSP cannot explain the observed photometry. The latter seems much more consistent with the mixed populations model, with a young mass fraction of $10^{-2}$ in the central region, which decreases towards $10^{-3}$ in the outskirts. No dust correction is needed to match the mixed populations tracks.
\subsection*{LARS02} 
\begin{figure*}
  \begin{center}
    \makebox[\textwidth]{\includegraphics[width=17cm]{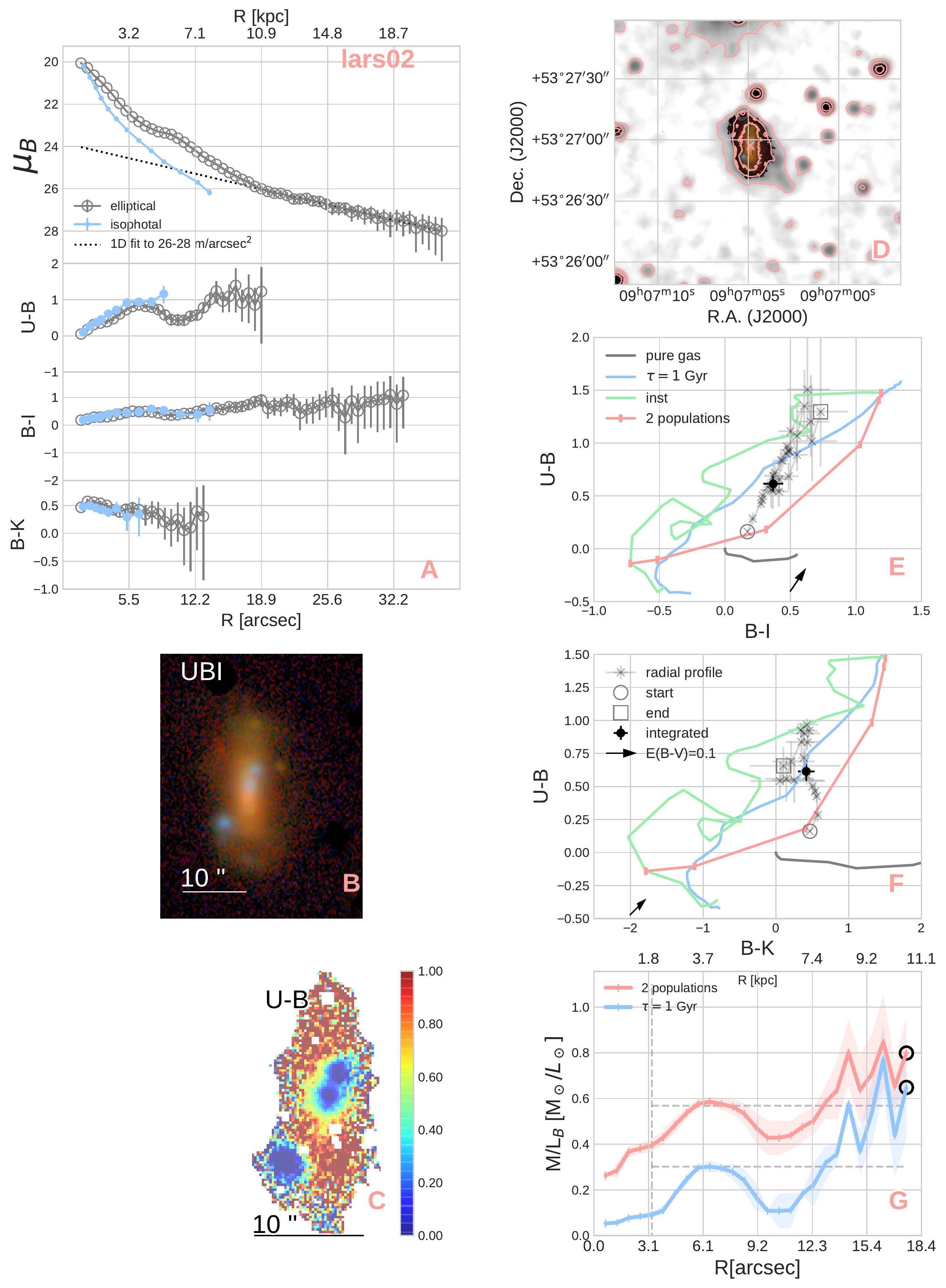}}
    \caption{Same as Figure~\ref{fig:lars01} but for LARS02.}\protect\label{fig:lars02}
  \end{center}
\end{figure*}

A star-forming region, offset from the geometric center, is evident in the radial profiles, RGB image, and $U-B$ spatial map. The three contours in panel D are at $24.5$, $26.2$, and $26.9$ mag/arcsec${}^2$, respectively. The faintest contour is irregular, hinting at a recent merger. Comparison to the SEMs in panels E and F in Fig. \ref{fig:lars02} shows that our mixture of two stellar populations with a young mass fraction of $10^{-2}$ is more consistent with the central regions of the galaxy, while the outskirts of the radial profiles are well-matched by a $1\mbox{-}2$ Gyr-old, metal poor stellar population. The latter is also a good match for the integrated color. The decrease in the M/L ratio around $\sim10$ arcsec corresponds to the offset star-forming region to the south-east from the center of integration (pink cross in panel D).
\subsection*{LARS03}
\begin{figure}
  \begin{center}
    \includegraphics[width=\hsize]{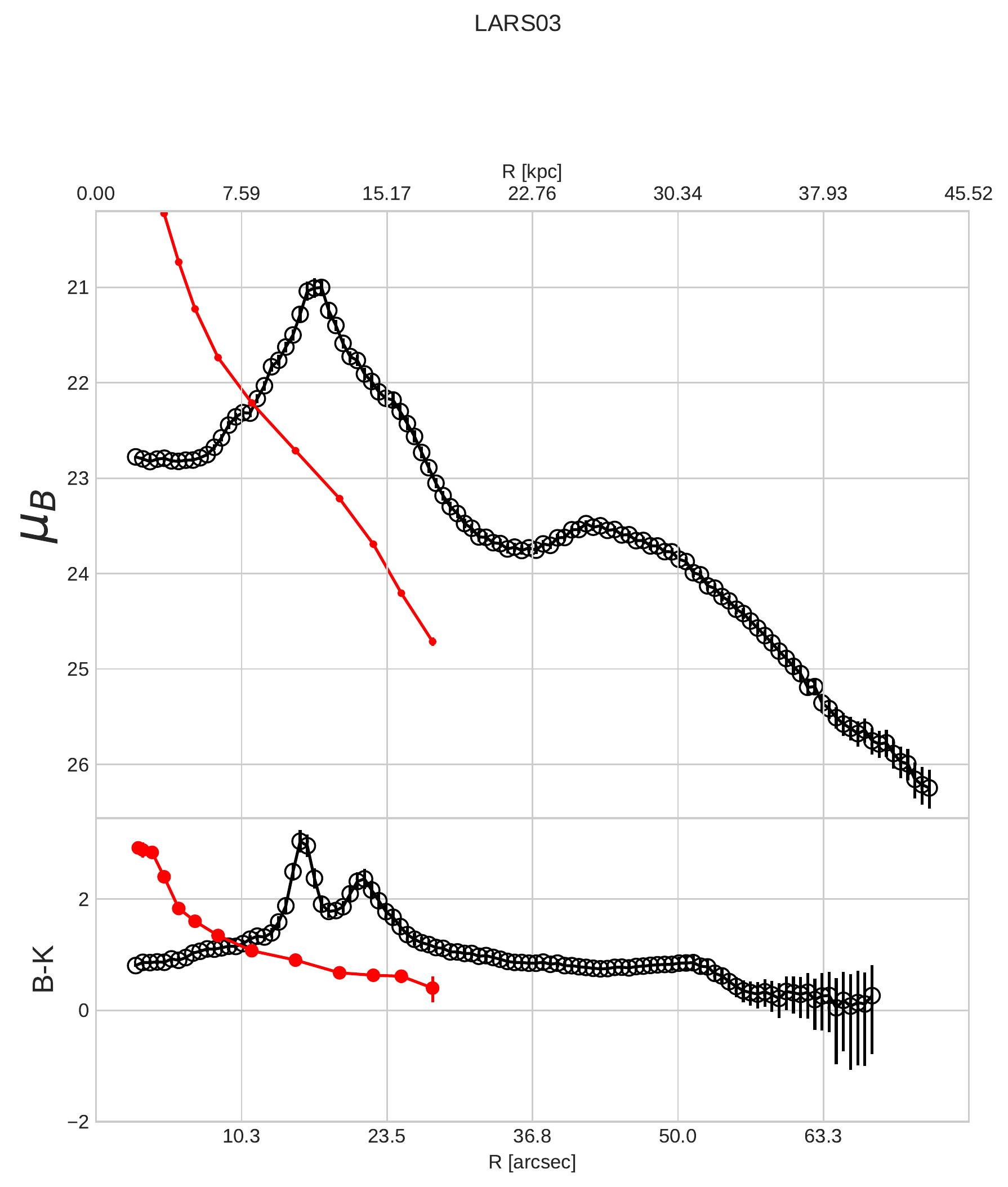}
    \includegraphics[width=\hsize]{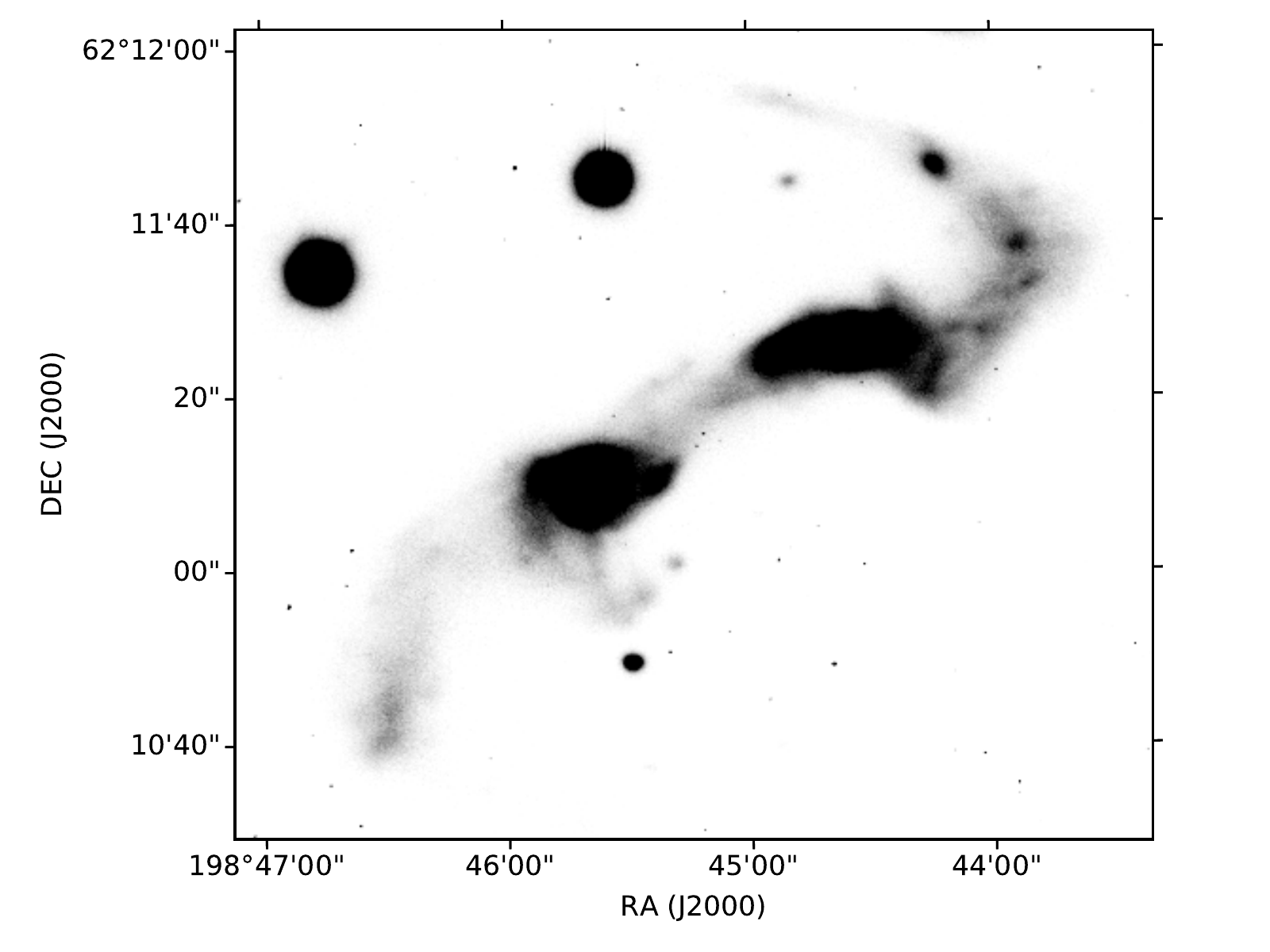}
    \caption{Top: $B$-band surface brightness and color profiles for LARS03. Bottom: $B$-band image of LARS03, demonstrating the complicated morphology of this galaxy. }\label{fig:lars03}
  \end{center}
\end{figure}

At the resolution of the ground-based data it is hard to separate the two merging galaxies in the LARS03 complex, which is why we selected not to complement the missing $U$ and $I$-bands for this galaxy with available {\it HST} imaging, which only covers the eastern part. Thus, only radial surface brightness and $B-K$ color profiles are presented for this galaxy in Fig.~\ref{fig:lars03}. Both indicate a complex structure.
\subsection*{LARS04}
\begin{figure*}
  \begin{center}
    \makebox[\textwidth]{\includegraphics[width=17cm]{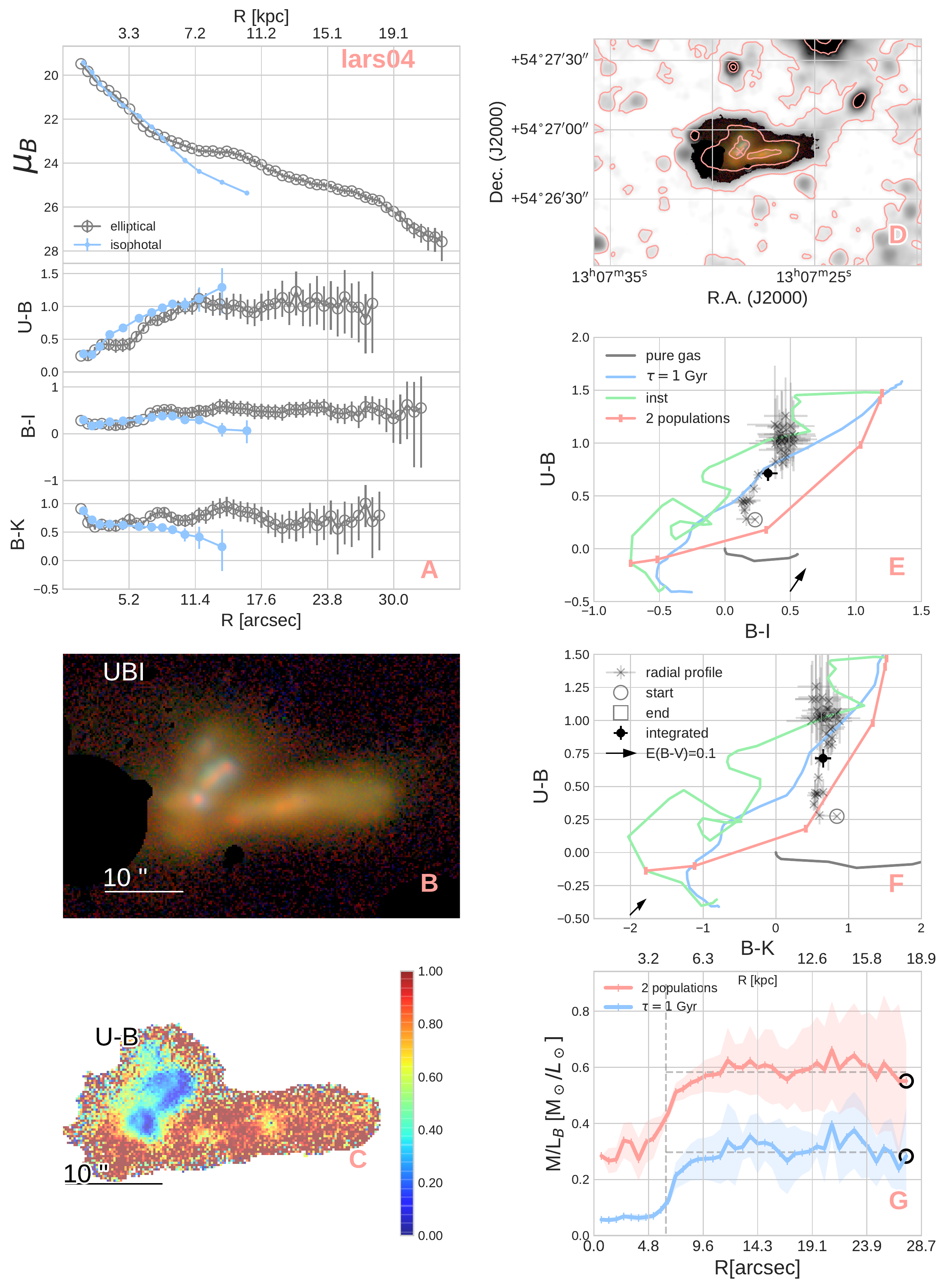}}
    \caption{As in Fig.~\ref{fig:lars01} but for LARS04.}\protect\label{fig:lars04}
  \end{center}
\end{figure*}

A star-forming region, offset from the geometric center, is evident in the radial profiles, RGB image, and $U-B$ spatial map. The three contours in panel D in Fig. \ref{fig:lars04} are at $21.9$, $24.9$, and $26.9$ mag/arcsec${}^2$, respectively, with the faintest contour tracing the shape of the brighter regions. This is clearly a merger. Similar to LARS02, the central region seems more consistent with a mixed populations SEM with a young mass fraction of $10^{-2}$, while the outskirts are well-matched by a $1\mbox{-}2$ Gyr-old, metal-poor, SSP model. The integrated color is consistent with a $1$ Gyr-old population.
\subsection*{LARS05}
\begin{figure*}
  \begin{center}
    \makebox[\textwidth]{\includegraphics[width=17cm]{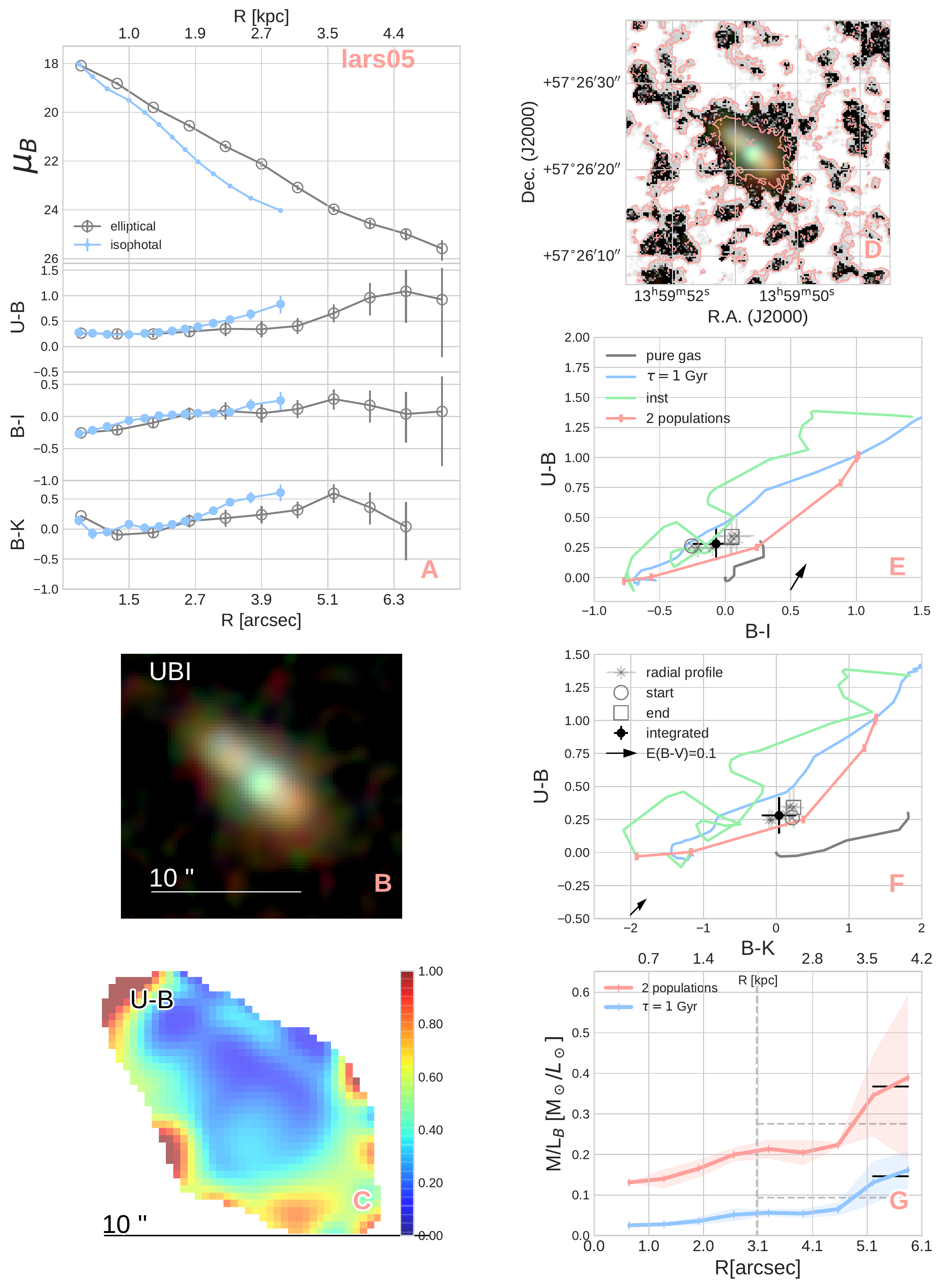}}
    \caption{As in Fig.~\ref{fig:lars01} but for LARS05.}\protect\label{fig:lars05}
  \end{center}
\end{figure*}

The two contours in panel D in Fig. \ref{fig:lars05} are at $23.4$ and $24.7$ mag/arcsec${}^2$. All optical data for this galaxy are {\it HST} data, resampled to the resolution of the $K$-band. For this reason, the faintest contour we could obtain is relatively bright and does not reveal any faint underlying structure. In the SEMs diagrams, the radial color profiles cluster around the same region, consistent with a mixed stellar populations model with a young mass fraction of $10^{-2}$. The figure suggests that the entire visible area of LARS05 in the optical data consists of well-mixed stellar populations, since no significant population change can be observed and the radial profile points cluster around similar values. 
\subsection*{LARS06}
\begin{figure*}
  \begin{center}
    \makebox[\textwidth]{\includegraphics[width=17cm]{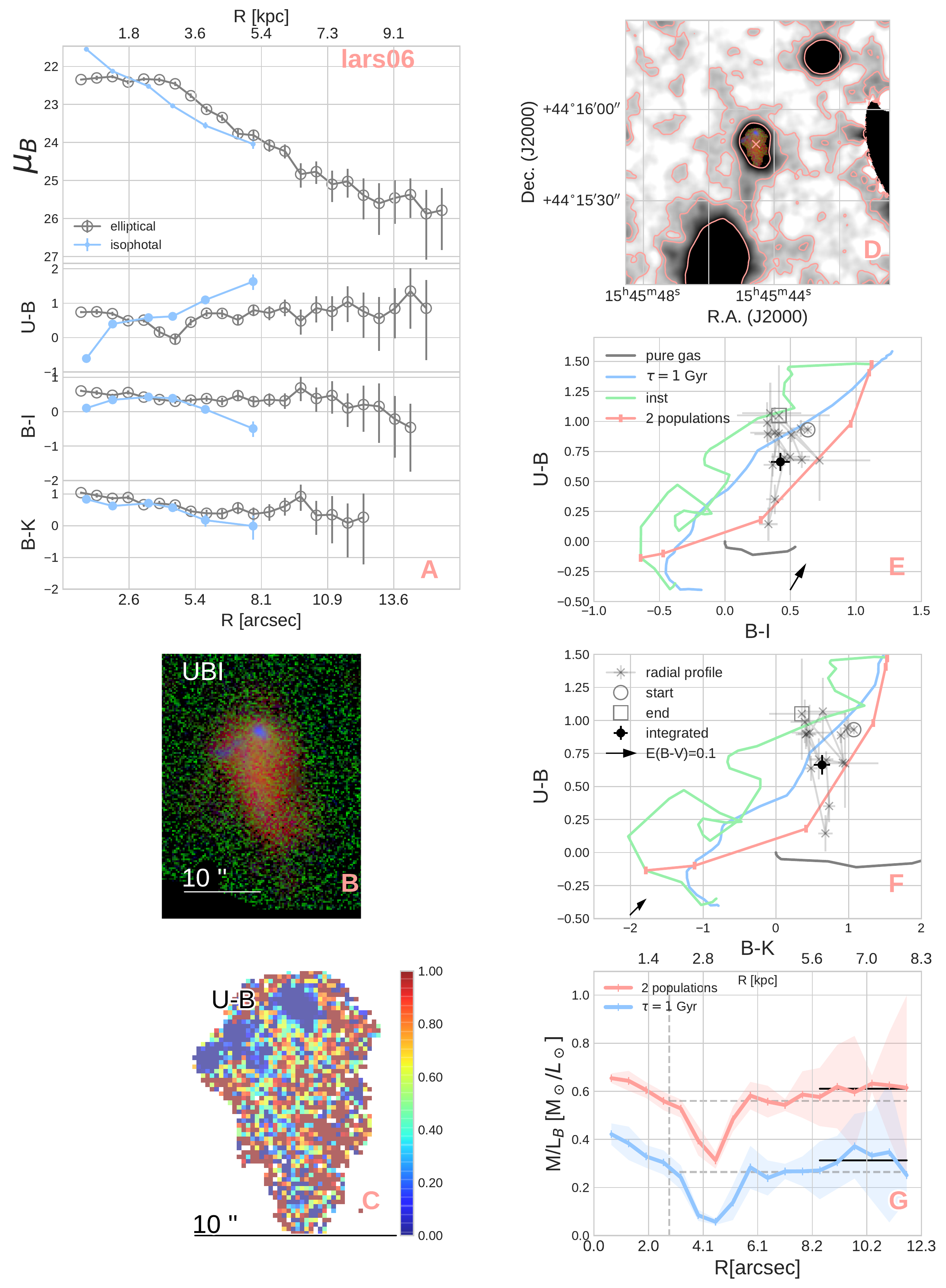}}
    \caption{As in Fig.~\ref{fig:lars01} but for LARS06.}\protect\label{fig:lars06}
  \end{center}
\end{figure*}

This is the faintest LARS galaxy, with several star-forming regions offset to the north and north-east from the geometric center. The three contours in panel D of Fig. \ref{fig:lars06} are at $23.9$, $24.3$, and $24.4$ mag/arcsec${}^2$, respectively. The faintest contour, albeit quite bright due to the low-luminosity nature of this galaxy, shows a very disturbed morphology, consistent with the presence of offset SF regions. We note that the center of integration (pink cross in panel D), is not on the star-forming region. This is reflected in the M/L ratio profile which decreases at $\sim4$ arcsec distance from the center. The SEMs also reflect this morphology, with both the central and outer radial color profiles being consistent with an old ($1.5$ Gyr) SSP model. In these color-color diagrams the radial color profiles at intermediate radial distances make an excursion in the direction of the two stellar populations model with a young mass fraction close to $10^{-2}$. As discussed in \citet{Pardy2014}, LARS06 may be interacting with the nearby field spiral UGC10028, located $\sim40$ arcsec to the south. These authors find that both galaxies are at the same velocity and the HI gas in LARS06 is significantly confused with the HI gas from its neighbor. Further supporting evidence for interaction between the two is the presence of an optical bridge at the $24.3$ mag/arcsec${}^2$ level, visible in panel D. The $3\sigma$ limiting surface brightness for this galaxy is the shallowest from all NOT data, $\mu_B^{lim}=26.1$ mag/arcsec${}^2$, preventing us from detecting the full extent of the optical bridge. 
\subsection*{LARS07}
\begin{figure*}
  \begin{center}
    \makebox[\textwidth]{\includegraphics[width=17cm]{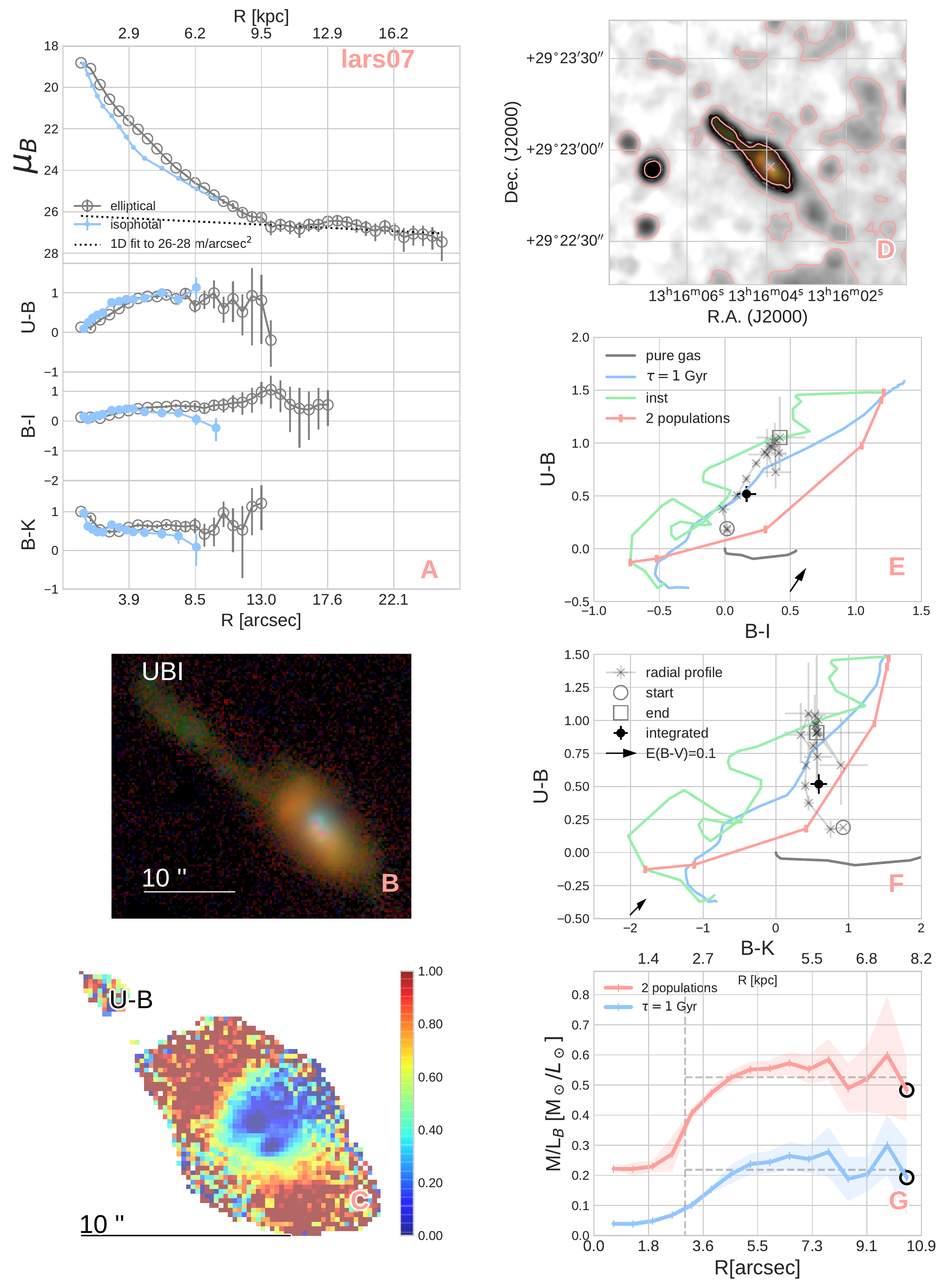}}
    \caption{As in Fig.~\ref{fig:lars01} but for LARS07.}\protect\label{fig:lars07}
  \end{center}
\end{figure*}

The central region of this galaxy shows an extended central SF region, surrounded by an older population, with the radial color profiles, RGB image, and $U-B$ map clearly reflecting such morphology. The two contours in panel D of Fig. \ref{fig:lars07} are at $24.8$ and $26.4$ mag/arcsec${}^2$. The faintest contour reveals two fainter substructures to the south-west, engulfed by the contour, and curiously aligned with the visible tail to the north-east of the central galaxy complex. It is possible that one or both of these substructures passed through the galaxy and are responsible for the observed morphology. The SEMs diagrams suggest a possible mixed population with a young mass fraction of $\sim10^{-2}$ towards the central regions, with the very center being more consistent with pure gas emission after correcting for dust attenuation, while the outskirts are consistent with an old ($\sim1.5$ Gyr), metal-poor, single stellar population. The integrated color, however, mimics a $400$-$500$ Myr single stellar population. The seeming decline of the M/L ratio at $\sim13$ arcsec is likely due to contribution to the $U-B$ and $B-I$ colors from the ``tail'' to the north-east. Both colors seemingly get bluer beyond this radius, suggesting the presence of young stars in the tail.
\subsection*{LARS08}
\begin{figure*}
  \begin{center}
    \makebox[\textwidth]{\includegraphics[width=17cm]{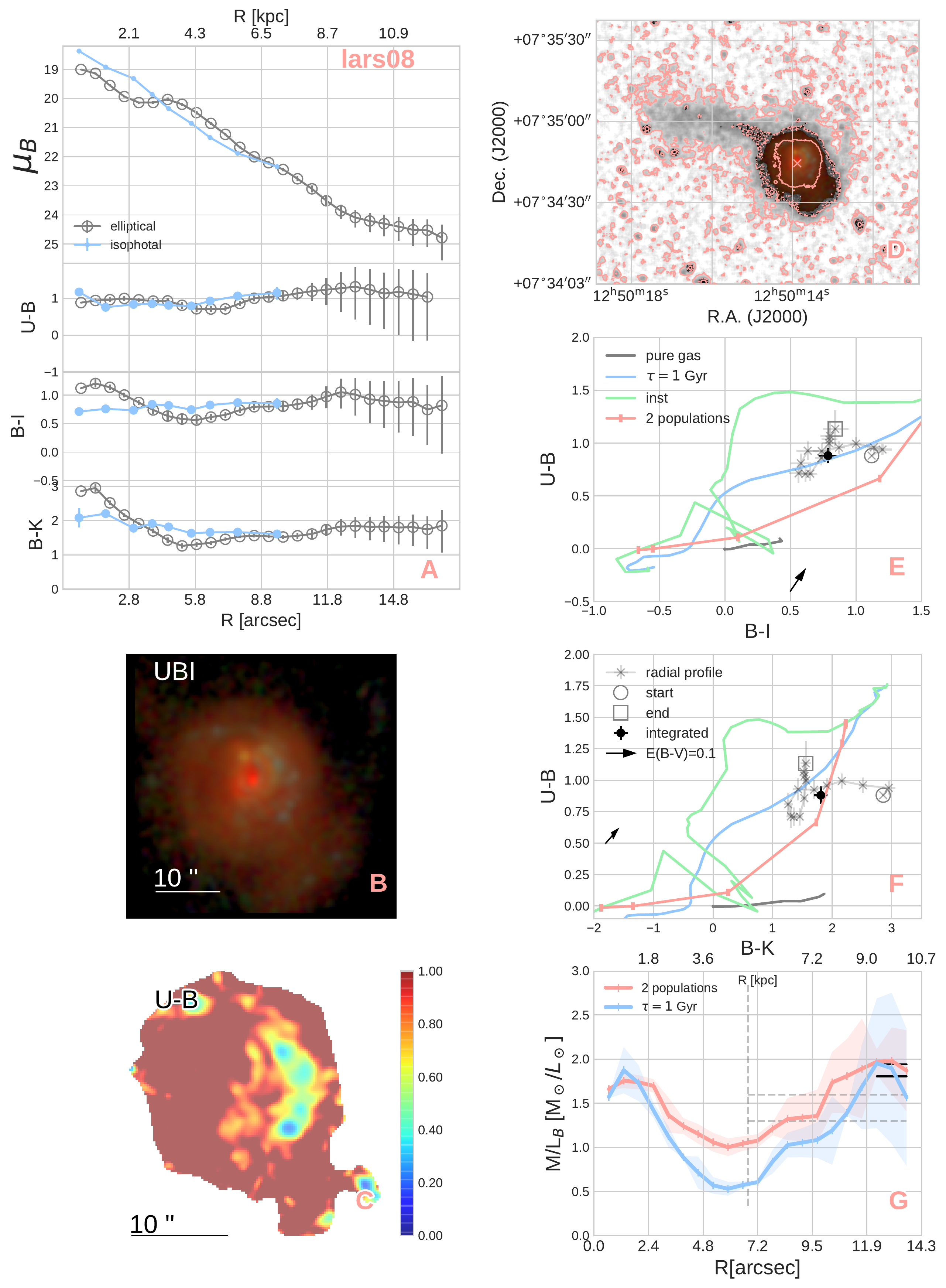}}
    \caption{As in Fig.~\ref{fig:lars01} but for LARS08.}\protect\label{fig:lars08}
  \end{center}
\end{figure*}

All optical data are {\it HST} images resampled to the $K$-band. The {\it HST} data covers only the central regions of the galaxy. In panel D of Fig. \ref{fig:lars08} we therefore show the $K$ image instead of a $B$-band image in order to show the tail to the north-east. The three $K$-band contours in panel D are at $20.8$, $22.8$, and $24.2$ mag/arcsec${}^2$. While the morphology is clearly disturbed, the radial color profiles, RGB image, and $U-B$ color maps show a seemingly older stellar population, assuming that the dust attenuation is negligible. However, in the presence of dust the central regions in panels F and G will move away from the SSP tracks towards the two population and gas tracks. The radial color profile and the integrated color occupy a model region of $2$-$3$ Gyr-old population. Interestingly, the $B-K$ versus $U-B$ diagram shows a central radial profile inconsistent with any of the three stellar tracks, even if we apply dust extinction correction along the plotted arrows. Only the pure gas track comes close to explaining these data points, after correction for dust attenuation. In this case the pure gas track is from the exponentially declining model, with $\tau=1$ Gyr. The M/L radial profile decreases at intermediate radii ($\sim5$ arcsec), reflecting the location of the SF regions. 
\subsection*{LARS09}
\begin{figure*}
  \begin{center}
    \makebox[\textwidth]{\includegraphics[width=17cm]{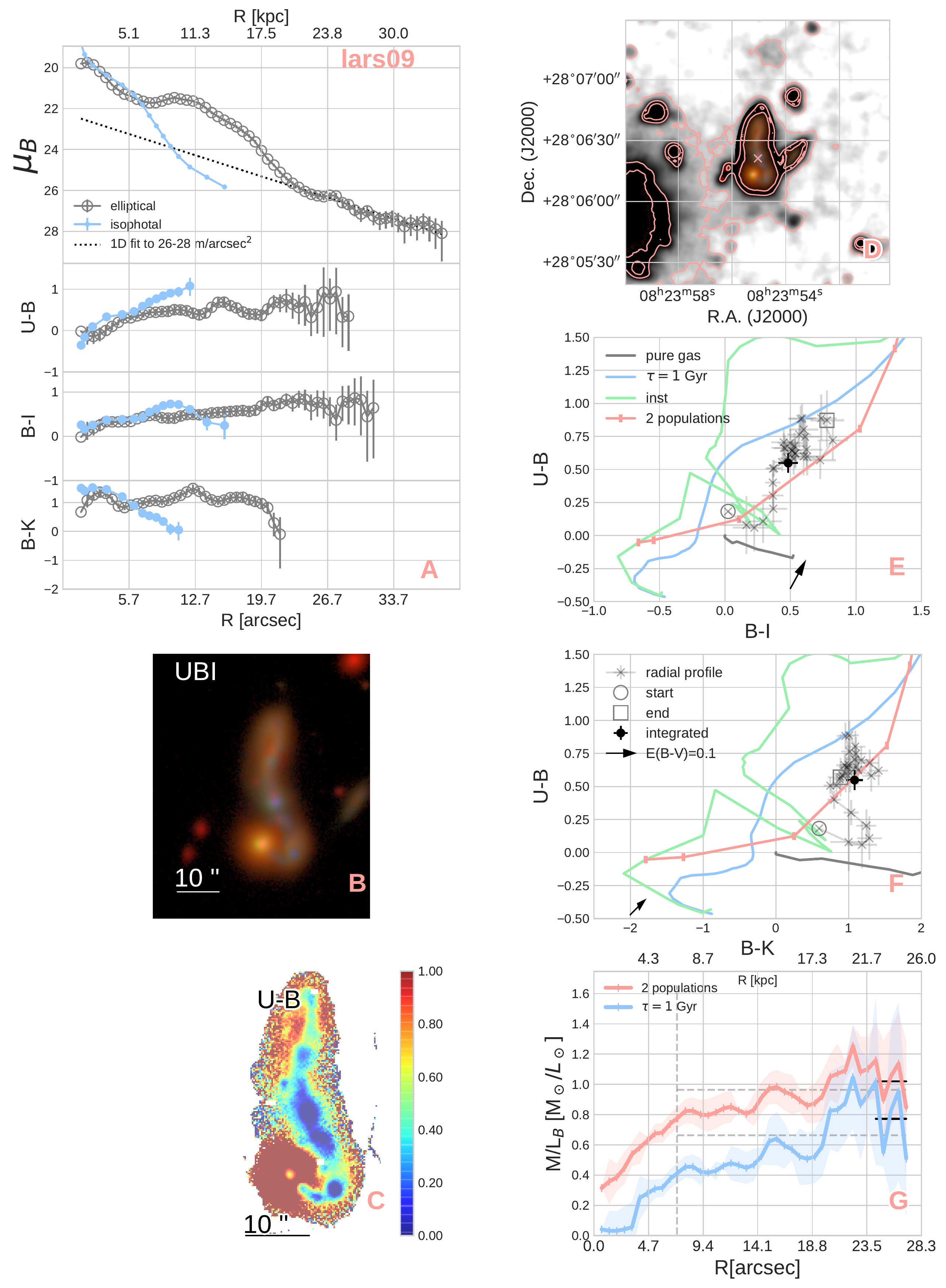}}
    \caption{As in Fig.~\ref{fig:lars01} but for LARS09.}\protect\label{fig:lars09}
  \end{center}
\end{figure*}

The bright star in the vicinity of LARS09 was masked out before any of the profiles were obtained. A complex, disturbed morphology with several SF regions is revealed by all analysis tools. The three contours in panel D of Fig. \ref{fig:lars09} are at $24.8$, $25.8$, and $26.8$ mag/arcsec${}^2$. The faintest contour engulfs a second galaxy to the west of LARS09. The RGB image shows that this galaxy has similar colors in UBI to LARS09, and is therefore unlikely to be a background galaxy. Indeed, H I data shows that this object is interacting with LARS09 (Cannon et al., in prep). The SEMs suggest that the central region of LARS09 is consistent with a mixed stellar population with a young mass fraction $10^{-2}$, which decreases to $\sim10^{-3}$ towards the outskirts, with some points deviating towards the pure gas track.
\subsection*{LARS10}
\begin{figure*}
  \begin{center}
    \makebox[\textwidth]{\includegraphics[width=17cm]{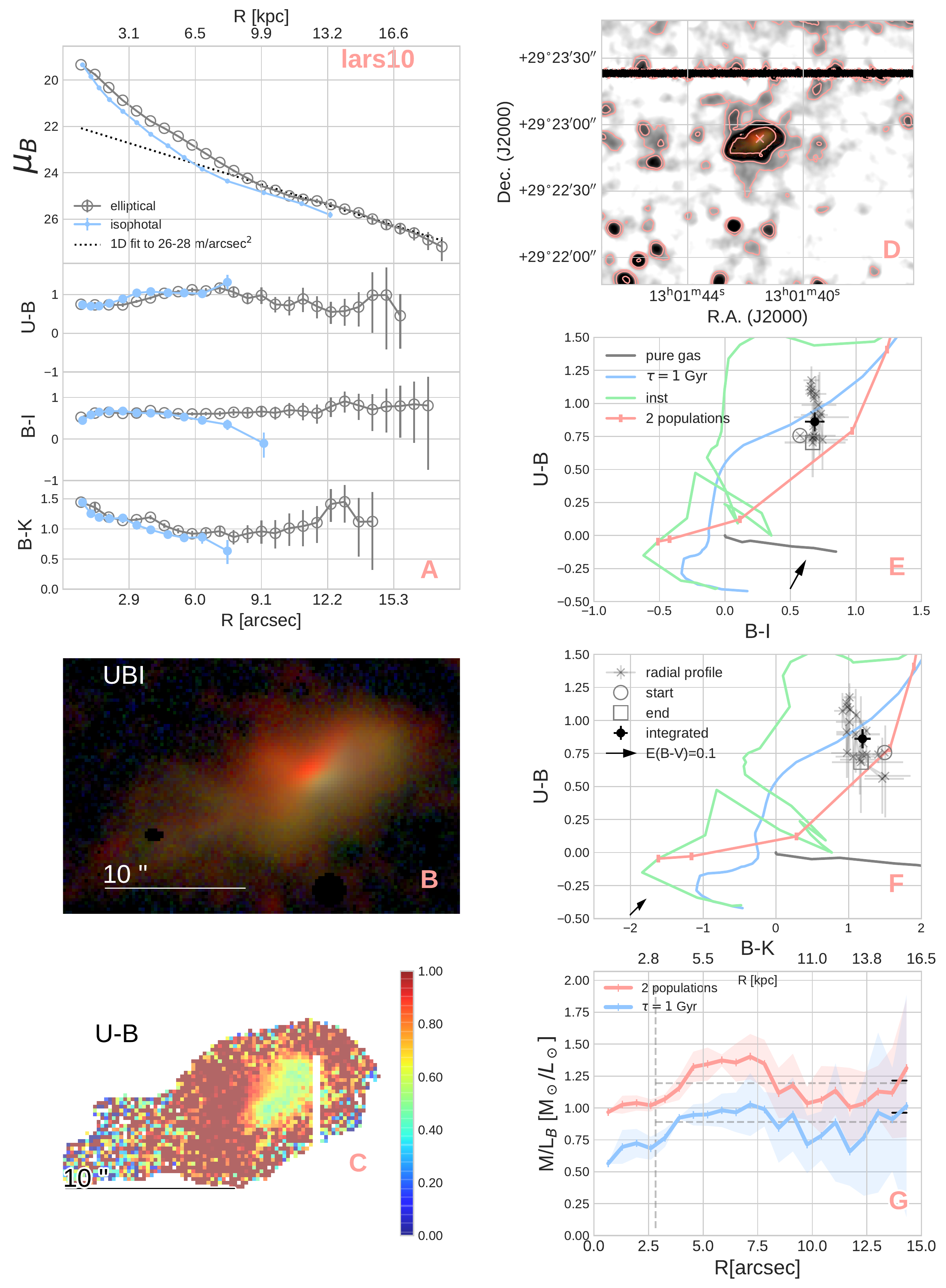}}
    \caption{As in Fig.~\ref{fig:lars01} but for LARS10.}\protect\label{fig:lars10}
  \end{center}
\end{figure*}

An extended central region with relatively blue colors is observed in the radial profiles, the RGB image, and the $U-B$ map. In both color-color panels, sections of the radial color profiles are only consistent with the two population mixture. The three contours in panel D of Fig. \ref{fig:lars10} are at $24.8$, $26.1$, and $26.3$ mag/arcsec${}^2$. The contour shows the irregular morphology of the faintest isophote we can reach, clearly suggesting a recent merger. At intermediate radii, a SSP model of solar metallicity and an age of $\sim2.5$ Gyr seems consistent with observations. The center, outskirts, and the integrated colors are more consistent with a mixed stellar population with a young mass fraction $\sim10^{-3}$.
\subsection*{LARS11}
\begin{figure*}
  \begin{center}
    \makebox[\textwidth]{\includegraphics[width=17cm]{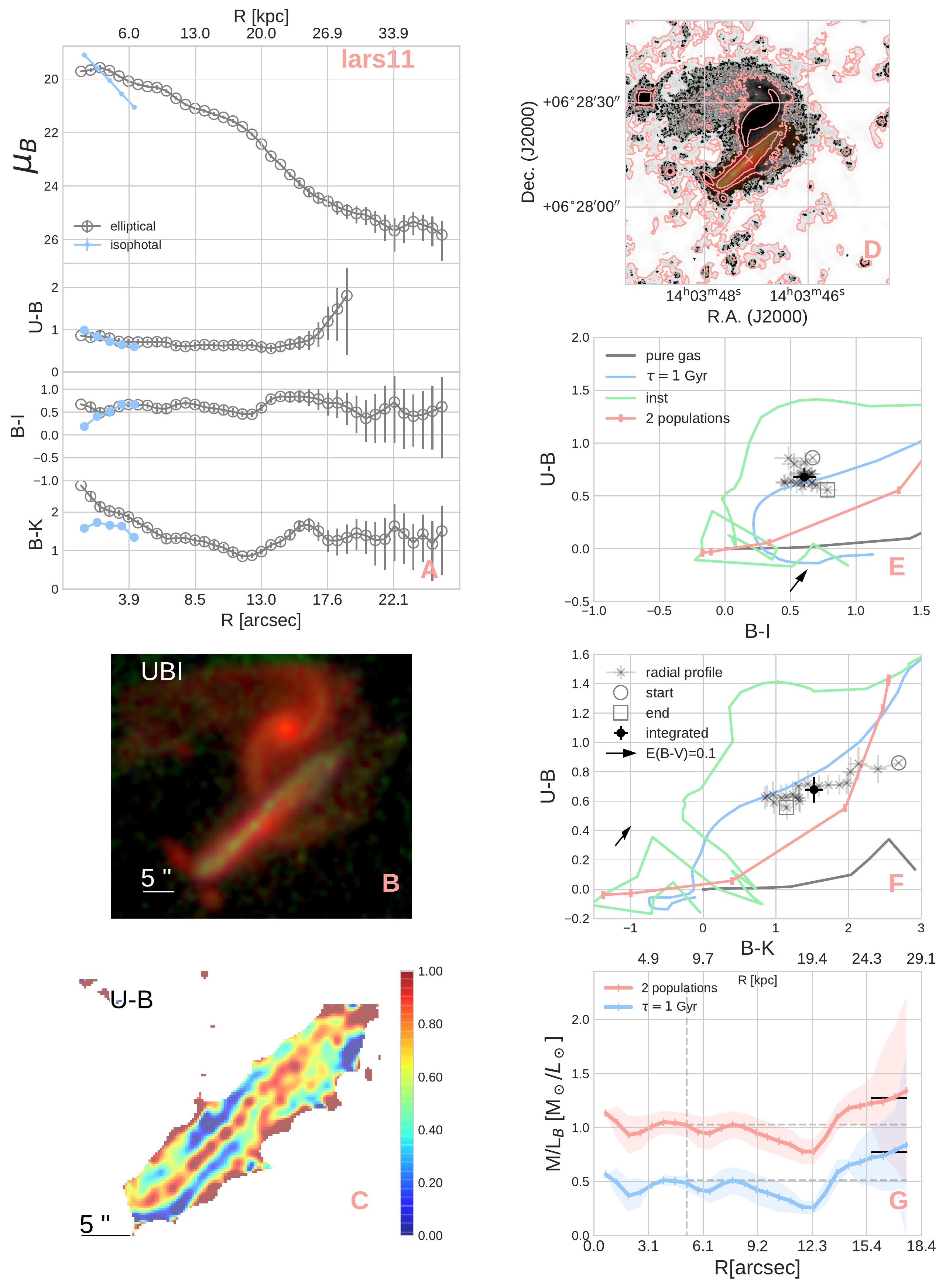}}
    \caption{As in Fig.~\ref{fig:lars01} but for LARS11.}\protect\label{fig:lars11}
  \end{center}
\end{figure*}

Due to the highly elliptical projection of this galaxy, the isophotal surface brightness profile starts to suffer from missing pixels in the outer isophotes very quickly. We are only confident in the first five radial steps of this profile, and omit the rest. All optical data are {\it HST} images resampled to $K$ resolution. The morphology is clearly disturbed, as seen in the radial profiles, RGB image, $U-B$ map, and contour plots. The three contours in panel D of Fig. \ref{fig:lars11} are at $20.6$, $22.1$, and $24.2$ mag/arcsec${}^2$. The spiral galaxy to the north of LARS11 is also clearly disturbed and it is difficult to confidently distinguish which debris belongs to LARS11. The two galaxies are interacting, as seen by CO tracing (Puschnig et al., in prep.). The SEM diagrams in panels E and F suggest that constant dust attenuation, applied to all points in the radial profile, will reconcile parts of the observations with the two population mixture with a young mass fraction between $10^{-2}\mbox{-}10^{-3}$, and with the pure gas emission track. The amount of attenuation depends on the assumed dust geometry. LARS11 has a nebular extinction of $0.477$ within the SDSS aperture and is thus the second most extinguished galaxy in LARS.
\subsection*{LARS12}
\begin{figure*}
  \begin{center}
    \makebox[\textwidth]{\includegraphics[width=17cm]{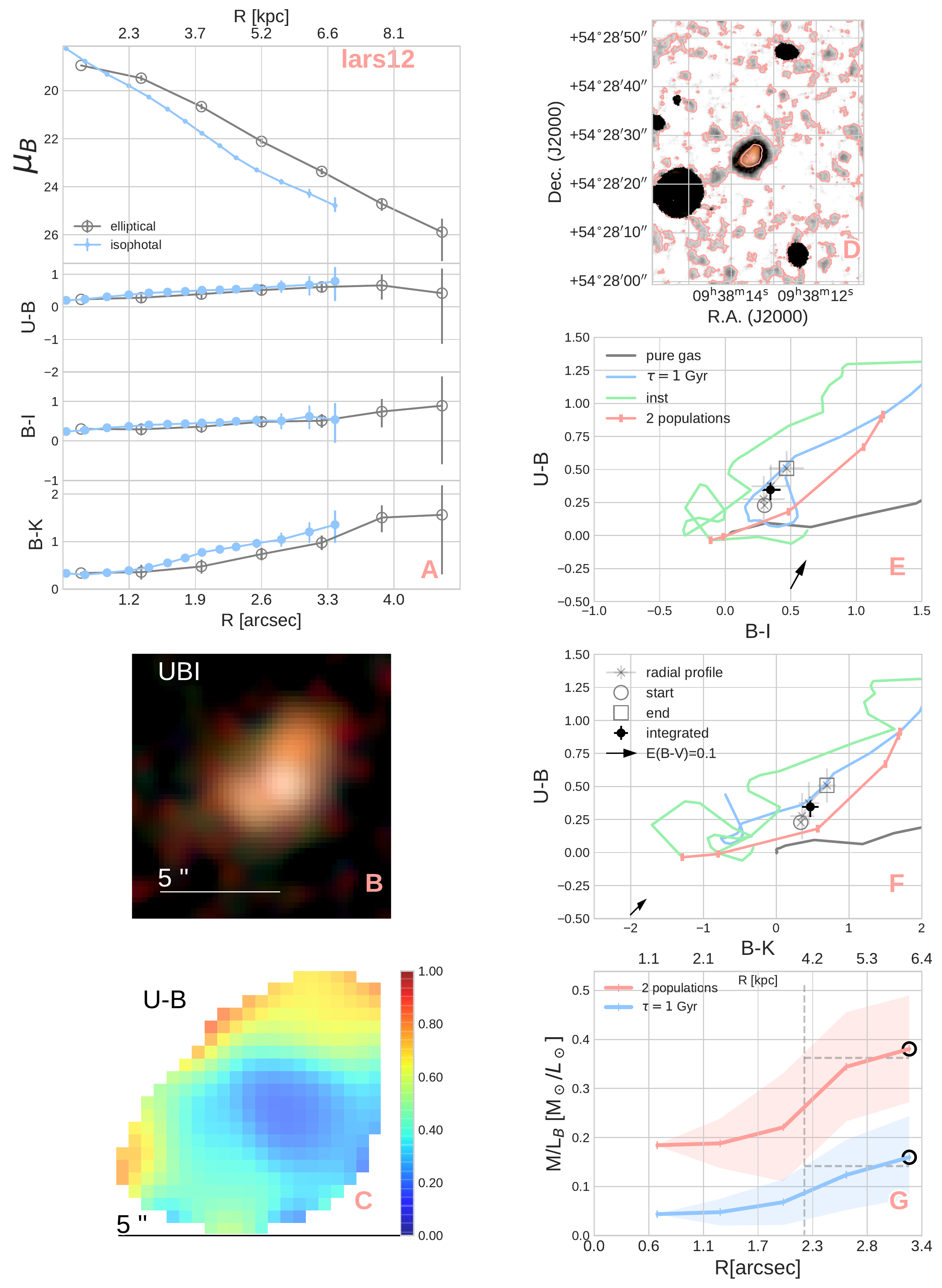}}
    \caption{As in Fig.~\ref{fig:lars01} but for LARS12.}\protect\label{fig:lars12}
  \end{center}
\end{figure*}

All optical data are {\it HST} images resampled to $K$ resolution. The radial color profiles are relatively flat, although B-K shows a clear trend of increasing domination of an older population. The $U-B$ map shows that the SF region seemingly dominates most of the galaxy visible in these data. The underlying gray scale image in panel D of Fig. \ref{fig:lars12} is the $K$-band because it is deeper than the {\it HST} optical data. The two contours are at $21.6$ and $24.4$ mag/arcsec${}^2$. The faintest contour in $K$ shows a fairly compact morphology, with a hint of debris to the north-west. The outskirts of the galaxy (last radial point) are consistent with a metal-poor SSP in the range $0.5\mbox{-}1$ Gyr, while the integrated color mimics a $300$ Myr-old population. The central region tends toward a mixed stellar population with a young mass fraction $>10^{-2}$.
\subsection*{LARS13}
\begin{figure*}
  \begin{center}
    \makebox[\textwidth]{\includegraphics[width=17cm]{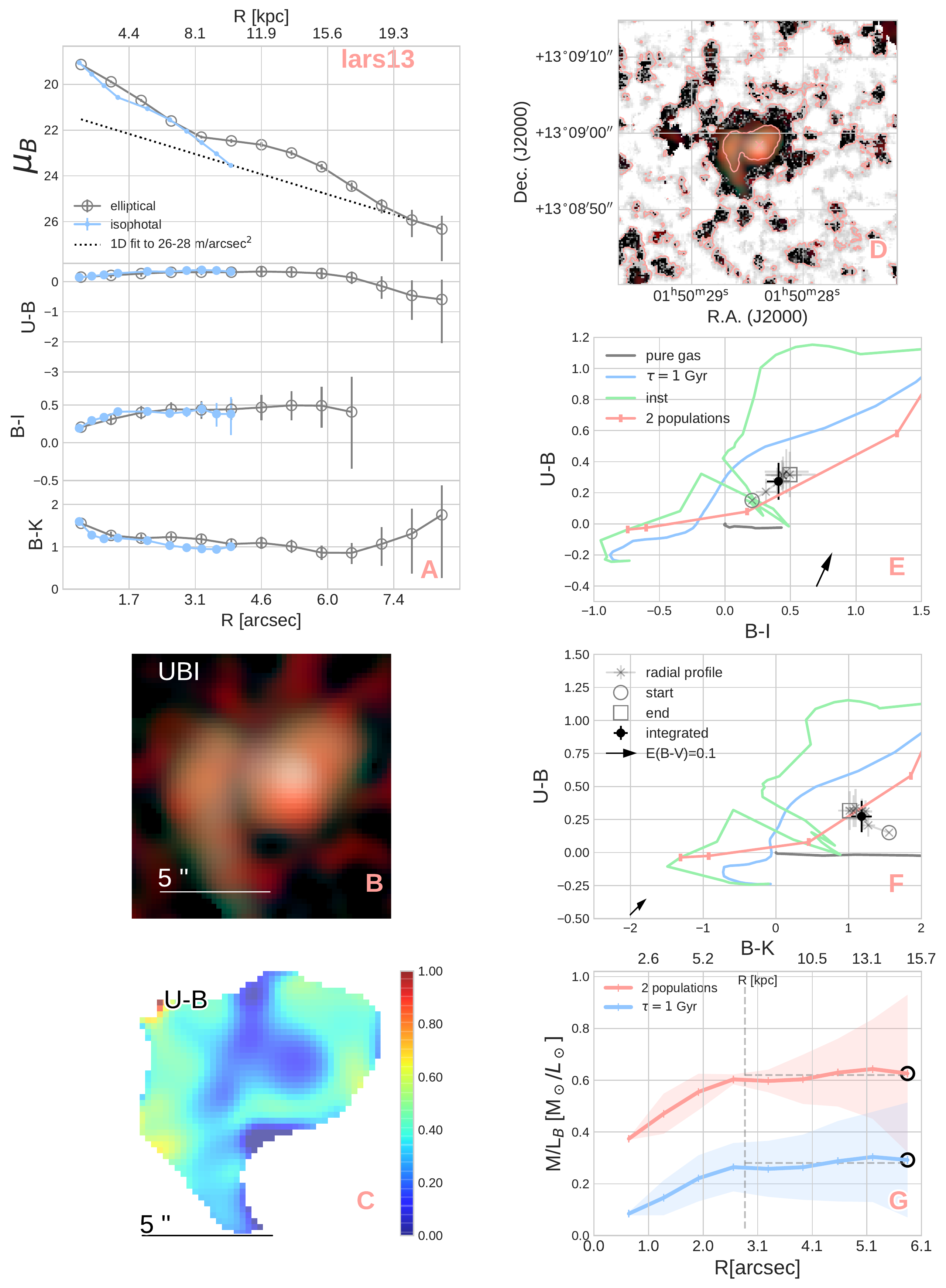}}
    \caption{As in Fig.~\ref{fig:lars01} but for LARS13.}\protect\label{fig:lars13}
  \end{center}
\end{figure*}

All optical data are {\it HST} images resampled to $K$ resolution. Similar to LARS12, this galaxy shows a compact morphology with a seemingly uniform single or well-mixed stellar population. Excluding the very center of the galaxy, the radial color profiles remain flat, indicating no change in population. This is also reflected in the flat M/L ratio profile. Panel D of Fig. \ref{fig:lars13} shows a gray scale $K$-band image, and the two contours are at $21.4$ and $24.1$ mag/arcsec${}^2$. The faintest contour indicates a clearly disturbed, complex morphology, indicating a merger event. The radial color profile data points are clustered in both SEM diagrams, supporting the lack of significant population change with radius. The two populations mixture with a young mass fraction between $10^{-2}\mbox{-}10^{-3}$ is more consistent with the outskirts of the radial profile and with the integrated color, while the central data point tends toward the pure gas emission track with negligible dust attenuation.
\subsection*{LARS14}
\begin{figure*}
  \begin{center}
    \makebox[\textwidth]{\includegraphics[width=17cm]{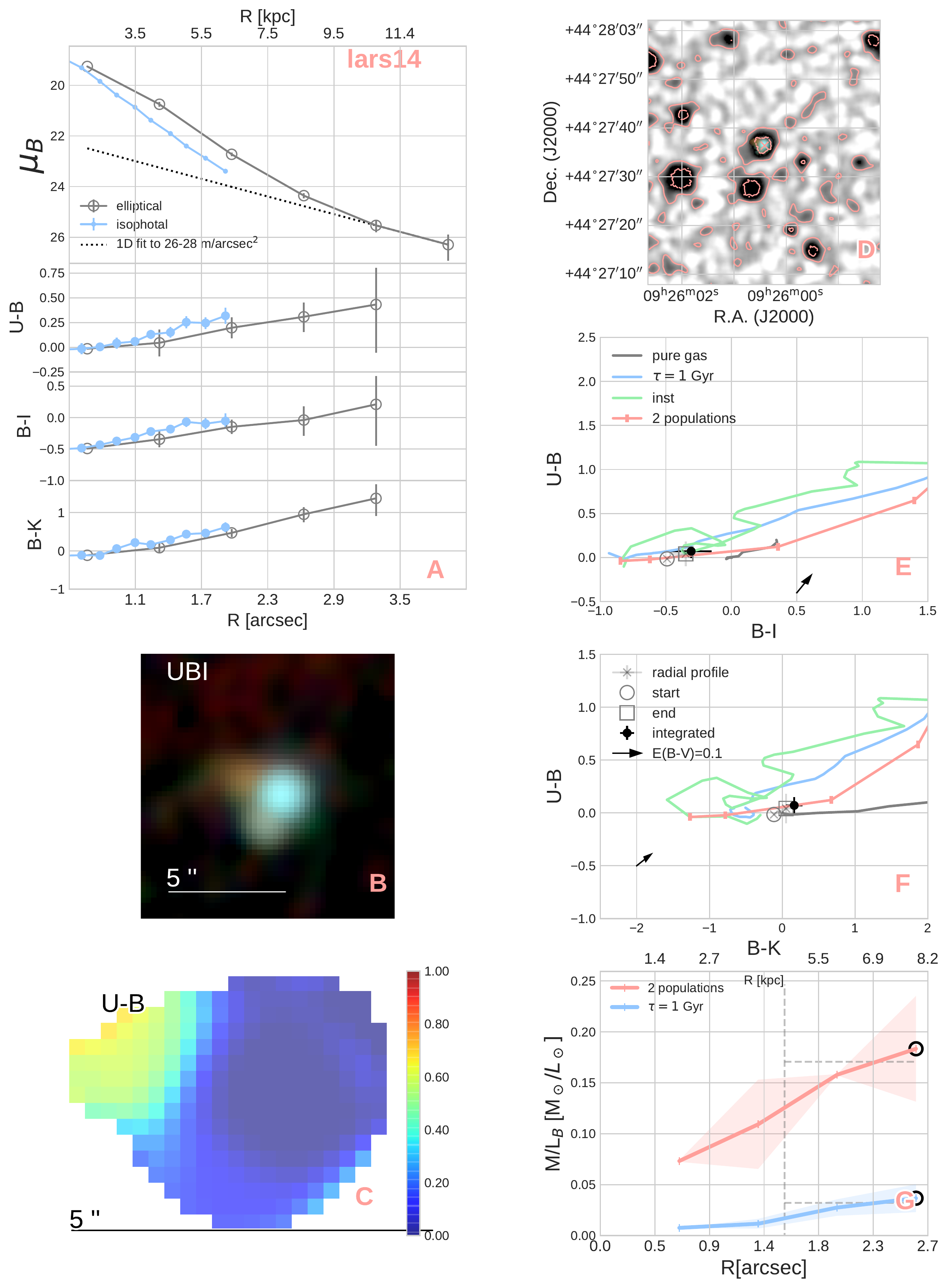}}
    \caption{As in Fig.~\ref{fig:lars01} but for LARS14.}\protect\label{fig:lars14}
  \end{center}
\end{figure*}

All optical data are {\it HST} images resampled to $K$ resolution. The radial color profiles monotonously redden in the outskirts, suggesting a progressively older population. The RGB image and the $U-B$ color map suggest the centeral regions are dominated either by a populaiton of a single age, or by a well-mixed distribution of several populations. This is supported by the clustering of the radial color profiles in the SEM diagrams. These points are consistent with the two populations mixture, with a young mass fraction higher than $10^{-2}$, with the center reaching $10^{-1}$, which is the highest in the sample. The gray-scale image in panel D of Fig. \ref{fig:lars14} is the $K$-band, and the two contours are at $22.3$ and $24.1$ mag/arcsec${}^2$. A compact morphology is revealed, although the extended structure to the north-west could be debris from a recent interaction.

\section*{Acknowledgments}

\noindent G.M. acknowledges support by the Swedish Research Council (Vetenskapsr\aa det). G.\"O. acknowledges support by VR and the Swedish National Space Board (SNSB). M.H. acknowledges the support of the Swedish Research Council, Vetenskapsr{\aa}det and the Swedish National Space Board (SNSB), and is Fellow of the Knut and Alice Wallenberg Foundation. D.K. is supported by the Centre National d’Etudes Spatiales (CNES)/Centre National de la Recherche Scientifique (CNRS); convention no 131425. \\

\noindent Based on observations made with the Nordic Optical Telescope, operated by the Nordic Optical Telescope Scientific Association at the Observatorio del Roque de los Muchachos, La Palma, Spain, of the Instituto de Astrofisica de Canarias.\\

\noindent The data presented here were obtained [in part] with ALFOSC, which is provided by the Instituto de Astrofisica de Andalucia (IAA) under a joint agreement with the University of Copenhagen and NOTSA.\\

\noindent Based [in part] on observations obtained with WIRCam, a joint project of CFHT, the Academia Sinica Institute of Astronomy and Astrophysics (ASIAA) in Taiwan, the Korea Astronomy and Space Science Institute (KASI) in Korea, Canada, France, and the Canada-France-Hawaii Telescope (CFHT) which is operated by the National Research Council (NRC) of Canada, the Institut National des Sciences de l'Univers of the Centre National de la Recherche Scientifique of France, and the University of Hawaii. \\

\noindent This research made use of Astropy, a community-developed core Python package for Astronomy \citep{AstropyCollaboration2013}.\\

\noindent This research made heavy use of the python packages \textsc{matplotlib} \citep{Hunter2007} and \textsc{seaborn}.\\

\noindent This research has made use of the NASA/IPAC Extragalactic Database (NED),
which is operated by the Jet Propulsion Laboratory, California Institute of Technology,
under contract with the National Aeronautics and Space Administration.\\

\noindent This research has made use of NASA's Astrophysics Data System Bibliographic Services (ADS).\\

\noindent \textit{Facilities}: NOT (MOSCA, ALFOSC), HST (ACS, WFC3), CFHT (WIRCAM)

\bibliographystyle{aa}
\bibliography{lars}

\label{lastpage}

\end{document}